\documentclass[]{interact}

\usepackage{epstopdf}
\usepackage[caption=false]{subfig}

\usepackage[numbers,sort&compress]{natbib}
\bibpunct[, ]{[}{]}{,}{n}{,}{,}
\renewcommand\bibfont{\fontsize{10}{12}\selectfont}

\theoremstyle{plain}

\theoremstyle{definition}

\theoremstyle{remark}

\usepackage{graphicx,epsfig}
\usepackage{latexsym}
\usepackage{amsfonts,amssymb}
\usepackage{amsmath}
\usepackage{natbib}
\usepackage{bm}
\usepackage{color}
\usepackage{xcolor}

\def \ii{{\mathrm{i}}}

\def \TL{{\mathrm{L}}}
\def \TP{{\mathrm{P}}}
\def \d{{\mathrm{d}}}

\def \pd{\partial}

\def \e{{\mathrm{e}}}

\def \tl#1{\overset{\kern 1pt\circ}{#1}}
\def \TL#1{\overset{\kern -3pt \circ}{#1}}
\def \TLL#1{\overset{\kern -7pt \circ}{#1}}

\def \Bbeta{\boldsymbol{\beta}}

\def \Bu{{\boldsymbol{u}}}

\def \Bx{{\boldsymbol{x}}}

\def \BR{{\boldsymbol{R}}}

\def \FF{{\cal{F}}}

\def \Bbeta{\boldsymbol{\beta}}

\def \Bu{{\boldsymbol{u}}}

\def\P{{\text P}}

\def\negenspace{\kern-1.1em}

\def\negenspaceexp{\kern-0.5em}

\begin{document}

\title{
A non-singular continuum theory of point defects using 
gradient elasticity of bi-Helmholtz type}
\author{
\name{Markus Lazar~$^\text{}$\footnote{
{\it E-mail address:} lazar@fkp.tu-darmstadt.de (M.~Lazar).}}
\affil{Department of Physics,
        Darmstadt University of Technology,
        Hochschulstr. 6,
        D-64289 Darmstadt, Germany}
}

\maketitle

\begin{abstract}
In this paper, 
we develop a non-singular continuum theory of point defects
based on a second strain gradient  elasticity theory, 
the so-called gradient elasticity of bi-Helmholtz type. 
Such a generalized continuum theory possesses a weak nonlocal character with two internal material lengths
and provides a mechanics of defects without singularities. 
Gradient elasticity of bi-Helmholtz type gives a natural and physical regularization of the classical 
singularities of defects, based on higher order partial differential equations. 
Point defects embedded in an isotropic solid are considered as eigenstrain problem in gradient elasticity of bi-Helmholtz type. 
Singularity-free fields of point defects are presented. 
The displacement field as well as the  first, the second and the third
gradients of the displacement are derived  
and it is shown that the classical singularities are regularized in this framework.  
This model delivers
non-singular expressions for the displacement field, the first
displacement gradient and the second displacement gradient.
Moreover, the plastic distortion (eigendistortion) and the gradient of the plastic distortion of a dilatation centre are also non-singular
and are given in terms of a form factor (shape function) of a point defect. 
Singularity-free expressions for the interaction energy and the interaction force between two dilatation centres 
and for the interaction energy and the interaction force of a dilatation centre in the stress field of an edge dislocation
are given.  
The results are applied to calculate the finite self-energy of a dilatation centre.
\end{abstract}


\section{Introduction}

Among the important problems widely addressed in the mechanics of defects 
is the determination of the elastic state of point defects. 
Defects (dislocations and point defects) are always present in crystalline solids. 
The interaction between defects can change considerably many properties of crystals.
In addition to dislocations, which are line defects, 
point defects are also important defects in crystals~\cite{LB,Teodosiu,Balluffi}. 
Point defects can exist in different configurations such as 
vacancies, interstitials, substitutional atoms, and foreign atoms. 
The fields of point defects play an important role in determining the physical properties of 
solids. 
They cause volume change and interact with dislocations 
if dislocations climb. 
Point defects play a major role 
in many physical problems such as  X-ray scattering, internal friction phenomena, aggregation of defects,
dislocation locking and various diffusion processes~\cite{Nabarro,LB}. 
Nowadays, the fields caused by point defects are important in
computer simulations of defect mechanics and discrete dislocation dynamics.
Atomistic and ab initio simulations of point defects and their interaction with
dislocations, including a comparison between atomic simulations and classical elasticity
theory, are reported in~\cite{Clouet06,Clouet13,Clouet18}.

Using the theory of classical elasticity, the displacement field and the elastic strain of point defects possess  
a $1/R^2$-singularity and a $1/R^3$-singularity, respectively~\cite{Teodosiu}.  
Moreover, the elastic strain of a point defect possesses a Dirac $\delta$-singularity~\cite{Eshelby55,Eshelby56,Lazar17}.
Such singularities are the result that classical elasticity
is not valid at the small scale near defects. Therefore, nonlocal elasticity or
strain gradient elasticity should be used
in order to obtain non-singular fields of point  defects in the close vicinity of the defect.

A particular version of Eringen's theory  of nonlocal elasticity~\cite{Eringen}
was used by~Kov\'acs~\cite{Kovacs81}, Wang~\cite{Wang} and Povstenko~\cite{Povstenko} to obtain non-singular stress fields whereas the displacement and elastic strain fields
still possess the classical singularities. 
Gairola~\cite{Gairola76,Gairola78} studied the elastic interaction between point defects using the nonlocal theory of elasticity.  
V{\"o}r{\"o}s and Kov\'acs~\cite{Kovacs93,Kovacs95} investigated the elastic interaction between point defects and dislocations in a Debye quasi-continuum. 

On the other hand, 
dilatation point defects (dilatation centres)
were investigated by Adler~\cite{Adler} 
in the framework of Toupin-Mindlin's first strain gradient 
elasticity~\cite{Toupin62,Mindlin64}.
It was shown by Adler~\cite{Adler} 
by using the linearized form of Toupin's strain-gradient theory~\cite{Toupin62}
that in an isotropic material the displacement field produced by a dilatation
point defect contains, besides the classical $1/R^2$-term, only exponentially
decreasing contributions and the expression given in~\cite{Adler} is 
singular. 
Later, in a simplified gradient elasticity theory, 
Dobov\v{s}eck~\cite{Dobov06} showed that the displacement field of a dilatation point defect
is non-singular and finite, but the strain and the strain gradient are still
singular, but weaker singular than the classical singularities. 
Vlasov~\cite{Vlasov} studied the elastic interaction between point defects and dislocations using a theory of 
simplified strain gradient elasticity. 
Theory of first strain gradient elasticity is unable to regularize and to remove all the classical singularities  
appearing in the elastic fields of point defects. Therefore, strain gradient elasticity of higher order is needed for a 
realistic modelling of point defects described by singularity-free elastic fields. 
Moreover, using a simple version of Toupin's theory of gradient elasticity at
finite strains with only one gradient term,
Wang et al.~\cite{Wang16} obtained numerical solutions of a dilatation centre 
reproducing a non-singular and finite
displacement field like in the linear theory of 
strain gradient elasticity obtained in~\cite{Dobov06}.

Furthermore, 
second strain gradient elasticity theory 
(or gradient elasticity of grade-3)
had originally been introduced
by Mindlin~\cite{Mindlin65,Mindlin72} (see also~\cite{Jaunzemis,Wu,AL09}). 
For the isotropic case, 
Mindlin's theory of second strain gradient elasticity 
involves sixteen additional material constants 
(including a so-called modulus of cohesion),
in addition to the two Lam\'e constants.
These material constants give rise to four characteristic length scales. 
Gradient elasticity of grade-$n$ can be considered as the continuum version of a lattice theory with $n$th-neighbour 
interactions~\cite{Mindlin65,Wu}.
In this way, Mindlin~\cite{Mindlin65} connected second strain gradient elasticity (gradient elasticity of grade-3) 
with  a one-dimensional lattice theory
with first, second and third neighbour interactions (see also~\cite{TV}).

Neglecting the modulus of cohesion and
the coupling terms between the strain tensor and the second strain gradient
tensor, 
a simplified and robust version of Mindlin's second strain gradient
elasticity, 
the so-called gradient elasticity theory of bi-Helmholtz type,
was introduced by Lazar et al.~\cite{LMA06} (see also~\cite{LM05,Poli03,Zhang})
and applied to problems of 
straight dislocations~\cite{LMA06,LM06}, 
dislocation loops~\cite{Lazar13},
straight disclinations~\cite{Deng2007} and cracks of 
mode I, II and III~\cite{Mousavi16-1,Mousavi16-2}.
Lazar et al.~\cite{LMA06,LM06} have shown that all physical state quantities
of screw and edge dislocations are non-singular in this framework.
Using gradient elasticity of bi-Helmholtz type,
it is possible to eliminate not only
the singularities of the strain and stress tensors, 
but also the singularities of the double and triple stress tensors and 
of the dislocation density tensors of straight dislocations.
All fields calculated in the theory of gradient elasticity of bi-Helmholtz type
are smoother than those calculated in gradient elasticity theory of Helmholtz
type.
The non-singular dislocation densities of screw and edge dislocations
obtained by~Lazar et al.~\cite{LM06,LMA06} in the framework of gradient elasticity
of bi-Helmholtz type 
have been recently used by Lyu and Li~\cite{Li19} in the multiscale modelling of
dislocation pattern dynamics.
Gradient elasticity of bi-Helmholtz type
is able to retain the main physical and mathematical features 
of Mindlin's second strain gradient elasticity. 
Moreover, gradient elasticity of bi-Helmholtz type corresponds to nonlocal elasticity of bi-Helmholtz
type with a nonlocal kernel  being the Green function of a differential operator of fourth order 
of bi-Helmholtz type~\cite{LMA06,LMA06b}. The Cauchy stress in gradient elasticity of bi-Helmholtz type and 
the stress in nonlocal elasticity of bi-Helmholtz type give coinciding and
non-singular solutions.
Under specific assumptions for the characteristic material lengths,
the dispersion relation of nonlocal elasticity of bi-Helmholtz type 
coincides with the  dispersion relation of the Born-K{\'a}rm{\'a}n model 
without lacking thermodynamic consistency~\cite{Eringen,LMA06,Fafalias}.

It should be noted that 
the influence of the coupling between the strain and the second strain gradient in Mindlin's strain gradient theory 
and a simplified version of it has been recently studied for one-dimensional problems of nano-objects in~\cite{Forest16,KN18}.

In this paper, we solve the long-outstanding problem of point defects in second strain gradient elasticity,
in order to find non-singular fields produced by point defects. 
We will answer the important mathematical question, up to which 
order $n$ of singularity $1/R^n$, strain gradient elasticity of bi-Helmholtz type 
is able to regularize the classical singularities and leading to singularity-free expressions in 3D.

The outline of this paper is as follows:
first, in Section~\ref{gradela} the theory of gradient
elasticity of bi-Helmholtz type, including Green functions, parameter study of the two appearing characteristic material lengths and operator split,
is presented.  
The problem of point defects is modelled as eigenstrain problem in second
strain gradient elasticity of bi-Helmholtz type.
In Section~\ref{PD}, the fields of a dilatation centre  
are computed and compared for gradient elasticity of bi-Helmholtz type, 
gradient elasticity of Helmholtz type and classical incompatible elasticity.
The interaction between two dilatation centres and the interaction between an edge dislocation and a dilatation centre
is studied in Section~\ref{Force}. 
We finally conclude in Section~\ref{Concl}.

\section{Gradient elasticity of bi-Helmholtz type}
\label{gradela}
 
 In this section, we consider the necessary basics, the three-dimensional Green tensor and the operator split of 
 the theory of incompatible gradient elasticity of bi-Helmholtz type
 originally proposed by Lazar et al.~\cite{LMA06}
 and we show that it is a suitable framework for a non-singular theory of point detects.
 
 \subsection{Basics of gradient elasticity of bi-Helmholtz type}
Consider  an infinite elastic body in a three-dimensional space  and  
assume that the  gradient of the displacement field $\bm u$ is additively
decomposed into an elastic distortion tensor  $\bm\beta$ and an inelastic\footnote{
The inelastic distortion comprises plastic and thermal effects, and is typically an incompatible field. When the inelastic distortion is absent the elastic distortion is compatible.
} distortion tensor (or eigendistortion tensor) $\bm \beta^\TP$:
\begin{align}
\label{uIJ}
\partial_ju_i=\beta_{ij}+\beta^\TP_{ij}\, .
\end{align}
For point defects, the eigendistortion tensor is often called quasi-plastic distortion tensor~\cite{Kroener60,deWit81}.  
The quasi-plastic distortion may serve as a defect density in its own right.  
As mentioned by Kr{\"o}ner~\cite{Kroener56,Kroener58,Kroener81} and 
deWit~\cite{deWit} the plastic distortion $\bm\beta^\TP$ of a point defect 
can be regarded as equivalent to  an ``infinitesimal dislocation loop density''
being a  fictitious dislocation distribution. 
The elastic strain tensor, $e_{ij}$, is the symmetric part of the elastic distortion tensor $\beta_{ij}$:
\begin{align}
\label{e-el}
e_{ij}=\frac{1}{2}\left(\beta_{ij}+\beta_{ji}\right)\, .
\end{align}

In the linear theory of gradient elasticity  of bi-Helmholtz type,
the strain energy density $\mathcal{W}$ 
of an isotropic material is given by~\cite{LM05,LMA06,Lazar13}
\begin{align}
\label{W-BH2}
\mathcal{W}=\frac{1}{2}\, C_{ijkl}e_{ij}e_{kl}
+\frac{1}{2}\, \ell_1^2 C_{ijkl}\pd_m e_{ij} \pd_m e_{kl}
+\frac{1}{2}\, \ell_2^4 C_{ijkl}\pd_n \pd_m e_{ij} \pd_n \pd_m e_{kl}
\, , 
\end{align}
where $\ell_1$ and  $\ell_2$ are  the two characteristic material lengths 
of gradient elasticity of bi-Helmholtz type
and $C_{ijkl}$ is the tensor of elastic moduli.
In gradient elasticity of bi-Helmholtz type, the strain energy density~\eqref{W-BH2} is given in terms of three contributions,
each one quadratic in elastic strain, first strain gradient and second strain gradient. 
Eq.~\eqref{W-BH2} can be derived from the form II of Mindlin's anisotropic second strain gradient elasticity theory
as a simplified version~\cite{LMA06}.
For isotropic materials, the tensor of elastic moduli 
reads
\begin{align}
\label{C}
C_{ijkl}=\lambda\, \delta_{ij}\delta_{kl}+
\mu\big(\delta_{ik}\delta_{jl}+\delta_{il}\delta_{jk}\big)\,,
\end{align}
where $\mu$ and $\lambda$ are the Lam\'e moduli.
The condition for positive semidefiniteness of the strain energy density, $\mathcal{W}\ge 0$, gives
\begin{align}
\label{pos-l}
\ell_1^2\ge 0\,,\qquad\ell_2^4\ge 0\,,
\end{align}
in addition to $(2\mu+3\lambda)\ge 0$ and $\mu\ge0$.
The positive semidefiniteness is a necessary condition for stable material
behaviour. 
If ${\cal W}$ is a quadratic form of its arguments, 
the strict convexity is
equivalent to the positive definiteness, ${\cal W}>0$
(see also~\cite{Forest16}).

The quantities canonically conjugated 
to the elastic strain tensor $e_{ij}$, 
the first strain gradient tensor (double strain) $\pd_ke_{ij}$
and the second strain gradient tensor (triple strain)  $\pd_l\pd_ke_{ij}$
are the Cauchy stress tensor $\sigma_{ij}$, the double stress tensor
$\tau_{ijk}$ and the triple stress tensor $\tau_{ijkl}$, respectively,  
and they give the following constitutive relations
\begin{align}
\label{CR1}
\sigma_{ij}&=\frac{\pd \mathcal{W}}{\pd e_{ij}}=C_{ijmn} e_{mn}\,,\\
\label{CR2}
\tau_{ijk}&=\frac{\pd \mathcal{W}}{\pd (\pd_k e_{ij})}
=\ell_1^2\,C_{ijmn} \pd_k e_{mn}=\ell_1^2 \pd_k \sigma_{ij}\,,\\
\label{CR3}
\tau_{ijkl}&=\frac{\pd \mathcal{W}}{\pd (\pd_l \pd_k e_{ij})}
=\ell_2^4\,C_{ijmn}\pd_l\pd_k e_{mn}=\ell_2^4\pd_l \pd_k \sigma_{ij}\,,
\end{align}
where we used that the elastic moduli are constant for homogeneous media. 
Eq.~\eqref{CR1} is nothing else than the Hooke law. 
It can be seen that $\ell_1$ is the characteristic length for 
the double stress tensor. 
On the other hand, $\ell_2$ is the characteristic length for 
the triple stress tensor.
It is remarkable, that the double stress tensor~(\ref{CR2}) is nothing but the first gradient of the stress tensor multiplied 
by $\ell_1^2$,
and that the triple stress tensor~(\ref{CR3}) is nothing but the second gradient of the stress tensor multiplied 
by $\ell_2^4$.
For a deeper discussion of the physical meaning of the double and triple
stresses we refer to~\cite{Poli16}.
Unlike Mindlin's second strain gradient elasticity~\cite{Mindlin65}, 
no coupling terms appear in the constitutive equations~\eqref{CR1}--\eqref{CR3} (see also \cite{LMA06}). 
However, a second strain gradient term in Eq.~\eqref{CR1} could give singular contributions to the 
Cauchy stress tensor $\sigma_{ij}$ which would not give a non-singular continuum theory of point defects.
A modulus of cohesion would just give a constant contribution to the triple stress tensor~\eqref{CR3}
being not relevant for the modelling of defects.

Using Eqs.~(\ref{CR1})--(\ref{CR3}), Eq.~(\ref{W-BH2}) 
can be re-written as~\cite{LMA06}
\begin{align}
\label{W-BH3}
\mathcal{W}=\frac{1}{2}\, \sigma_{ij}e_{ij}
+\frac{1}{2}\, \ell_1^2 \pd_k \sigma_{ij} \pd_k e_{ij}
+\frac{1}{2}\, \ell_2^4 \pd_l \pd_k \sigma_{ij} \pd_l \pd_k e_{ij}
\, .
\end{align}
The strain energy density~(\ref{W-BH3}) exhibits a stress-strain symmetry in
$\sigma_{ij}$ and $e_{ij}$, in $\pd_k \sigma_{ij}$ and $\pd_k e_{ij}$, 
and in $\pd_l \pd_k \sigma_{ij}$ and $\pd_l \pd_k e_{ij}$ (see also~\cite{LM05}).
Using the inverse tensor of the elastic moduli $C_{ijkl}^{-1}$,
the energy density (\ref{W-BH2}) can be written in an equivalent stress gradient form 
\begin{align}
\label{W-BH4}
\mathcal{W}=\frac{1}{2}\, C^{-1}_{ijkl}\sigma_{ij}\sigma_{kl}
+\frac{1}{2}\, \ell_1^2 C^{-1}_{ijkl}\pd_m \sigma_{ij} \pd_m \sigma_{kl}
+\frac{1}{2}\, \ell_2^4 C^{-1}_{ijkl}\pd_n \pd_m \sigma_{ij} \pd_n \pd_m \sigma_{kl}
\, .
\end{align}

An interesting side-remark,
using Eqs.~(\ref{W-BH3}) and (\ref{W-BH4}),
we may define the quantities canonically conjugated to the Cauchy stress tensor,
the first stress gradient tensor and the second  stress gradient tensor 
which are the tensors $e_{ij}$, $\phi_{ijk}$ and $\phi_{ijkl}$, respectively, 
and they give the following inverse constitutive relations
\begin{align}
\label{CR1-2}
e_{ij}&=\frac{\pd \mathcal{W}}{\pd \sigma_{ij}}=C^{-1}_{ijmn} \sigma_{mn}\,,\\
\label{CR2-2}
\phi_{ijk}&=\frac{\pd \mathcal{W}}{\pd (\pd_k \sigma_{ij})}
=\ell_1^2\,C^{-1}_{ijmn} \pd_k \sigma_{mn}=\ell_1^2 \pd_k e_{ij}\,,\\
\label{CR3-2}
\phi_{ijkl}&=\frac{\pd \mathcal{W}}{\pd (\pd_l \pd_k \sigma_{ij})}
=\ell_2^4\,C^{-1}_{ijmn}\pd_l\pd_k \sigma_{mn}=\ell_2^4\pd_l \pd_k e_{ij}\,.
\end{align}
Eq.~\eqref{CR1-2} is simply the inverse Hooke law. 
It is remarkable that $\phi_{ijk}$ is nothing but the first gradient of the
elastic strain tensor multiplied  by  $\ell_1^2$,
and $\phi_{ijkl}$ is nothing but the second gradient 
of the  elastic strain tensor multiplied by  $\ell_2^4$.
Since the strain $e_{ij}$ is dimensionless, $\phi_{ijk}$ 
has the dimension: [length], and $\phi_{ijkl}$ has the dimension: [length]${}^2$.
Such a tensor $\phi_{ijk}$ was originally introduced by Forest and Sab~\cite{Forest12} as micro-displacement tensor 
in stress gradient continuum theory (see also~\cite{Poli18}).
\\
It is interesting to note that 
the symmetry between the constitutive relations (\ref{CR1})--(\ref{CR3}) 
and the inverse constitutive relations~(\ref{CR1-2})--(\ref{CR3-2}) 
is based on the above mentioned symmetry between strains and stresses in
the strain energy density (\ref{W-BH3}) of gradient elasticity of bi-Helmholtz type.

In the presence of body forces $\bm b$, 
the  static Lagrangian density reads 
\begin{align}
\mathcal{L}&=-\mathcal{W}-\mathcal{V}\nonumber\,,
\end{align}
where 
\begin{align}
\label{V}
\mathcal{V}=-u_i b_i
\end{align}
is the potential of body forces.
The condition of static equilibrium  is expressed by the Euler-Lagrange equations.
In second strain gradient elasticity 
(linear elasticity of grade-3), 
the Euler-Lagrange equations read  (e.g., \cite{AL09})
\begin{align}
\label{EL-u}
\frac{\delta \mathcal{L}}{\delta u_i}=\frac{\pd \mathcal{L}}{\pd u_i}
-\pd_j\, \frac{\pd \mathcal{L}}{\pd (\pd_j u_i)}
+\pd_k \pd_j\, \frac{\pd \mathcal{L}}{\pd (\pd_k \pd_j u_i)}
-\pd_l \pd_k \pd_j\, \frac{\pd \mathcal{L}}{\pd (\pd_l\pd_k \pd_j u_i)}
=0\,.
\end{align}
In terms of the Cauchy stress, double stress and triple stress tensors, 
Eq.~(\ref{EL-u}) takes the form \cite{Mindlin65}
\begin{align}
\label{Eq-BH}
\pd_j(\sigma_{ij}-\pd_k \tau_{ijk}+\pd_l\pd_k \tau_{ijkl})+b_i=0\,.
\end{align}
Using Eqs.~(\ref{CR2}) and (\ref{CR3}), Eq.~(\ref{Eq-BH}) 
reduces to 
\begin{align}
\label{T-stress-BH2}
L \,\pd_j \sigma_{ij}+b_i=0\,,
\end{align}
where the fourth order differential operator $L$ is given 
by
\begin{align}
\label{L-BH}
L=\big(1-\ell_1^2\Delta+\ell_2^4\Delta\Delta\big)\,,
\end{align}
where $\Delta$ is the Laplace operator.

Substituting the constitutive relation~(\ref{CR1}) and 
Eq.~(\ref{uIJ}) into the equilibrium condition~(\ref{T-stress-BH2}), 
we obtain a partial differential equation of sixth order,
which is called bi-Helmholtz-Navier equation~\cite{LMA06,Lazar13},
for the displacement vector $\Bu$
\begin{align}
\label{u-L0}
L L_{ik} u_k =C_{ijkl} \pd_j L  \beta^{\TP}_{kl}-b_i\,,
\end{align}
where 
\begin{align}
\label{L-Navier}
L_{ik}=C_{ijkl}\pd_j\pd_l
\end{align}
is the Navier differential operator. 
For an isotropic material, it reads
\begin{align}
L_{ik}=\mu\, \delta_{ik}\Delta+(\mu+\lambda)\, \pd_i\pd_k\,.
\end{align}
Eq.~(\ref{u-L0}) is nothing but the equilibrium condition~(\ref{T-stress-BH2})
written in terms of the displacement vector $\Bu$ and the plastic distortion
tensor $\Bbeta^\TP$,
and
\begin{align}
\label{effectiveF}
f_i=b_i-C_{ijkl}\partial_jL\beta^\TP_{kl}
\end{align}
is the body force-type term. Notice that the second term on the 
right hand side  of Eq.~(\ref{effectiveF}) is an
``effective" or ``fictitious'' internal body force due to the 
gradient of the eigendistortion, 
and arises in the presence of defects, such as
point defects, dislocations, and disclinations. 
This term is the gradient version of  the internal force 
in Mura's eigenstrain method~\cite{Mura}.
For defects, the body force density vanishes, $b_i=0$.

Alternatively, the differential operator of fourth order~\eqref{L-BH}  can be written in the form as  product of two Helmholtz operators
(differential operators of second order)
\begin{align}
\label{L-BH-2}
L=\big(1-c_1^2\Delta\big)\big(1-c_2^2\Delta\big)
\end{align}
with
\begin{align}
\label{c1c2-1}
\ell_1^{2}&=c_1^{2}+c_2^{2}\, ,\\
\label{c1c2-2}
\ell_2^{4}&=c_1^{2}\, c_2^{2}\,
\end{align}
and 
\begin{align}
\label{c1-2}
c^{2}_{1,2}&=\frac{\ell_1^{2}}{2}\Bigg(1\pm\sqrt{1-4\,\frac{\ell_2^{4}}{\ell_1^{4}}}\Bigg)\,.
\end{align}
Here $c_1$ and $c_2$ are the two auxiliary lengths 
appearing in the two Helmholtz operators in Eq.~\eqref{L-BH-2}.  
Due to its structure as a product of two Helmholtz operators,
the differential operator~(\ref{L-BH-2}) is called bi-Helmholtz operator~\cite{LMA06,LMA06b}.

\begin{figure}[t]\unitlength1cm
\vspace*{0.1cm}
\centerline{
\epsfig{figure=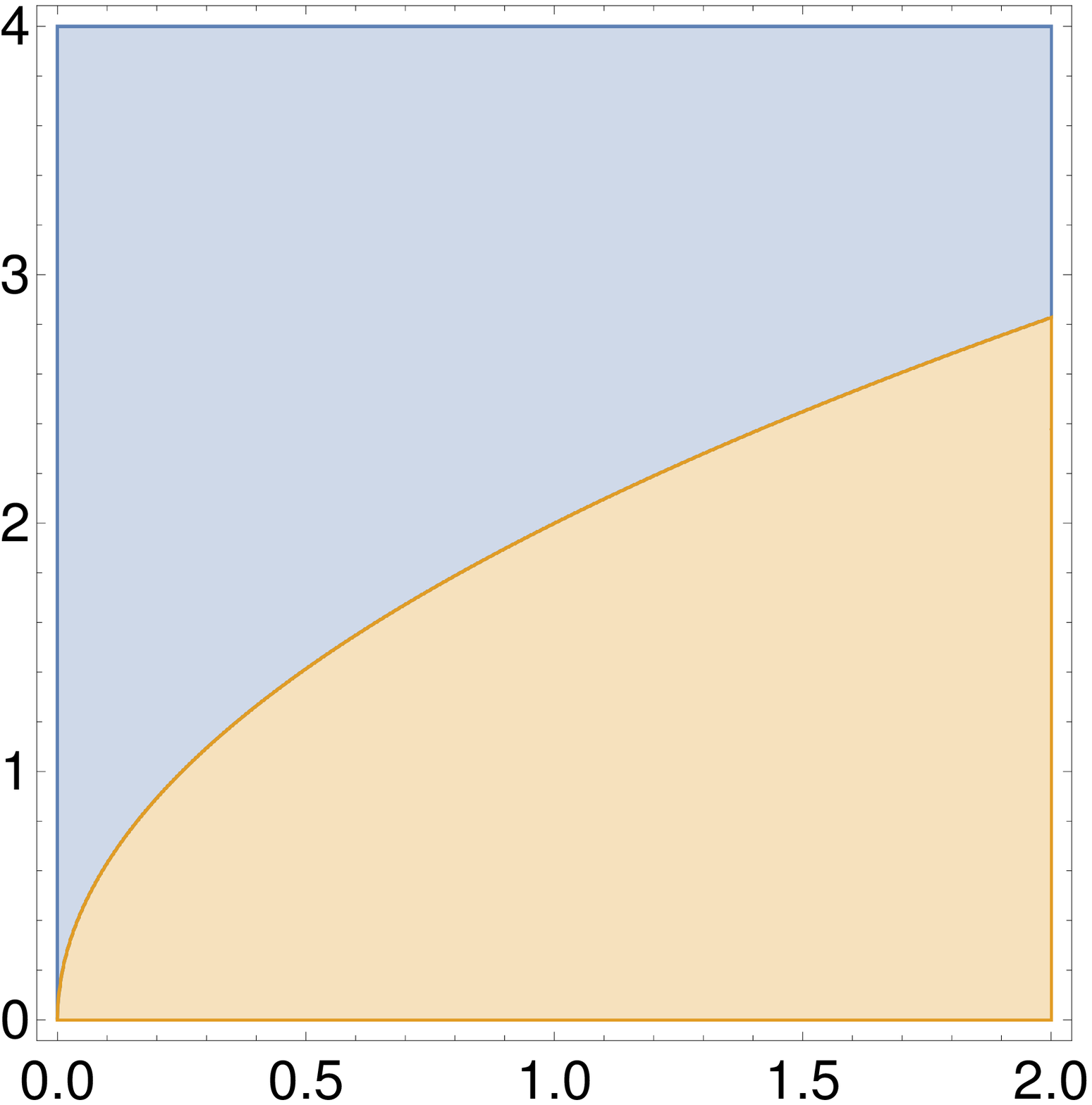,width=6.0cm}
\put(-3.1,-0.6){$\ell_2^4$}
\put(-6.6,3.0){$\ell_1^2$}
\put(-1.5,1.0){$\text{\boxed{3}}$}
\put(-2.5,3.3){$\text{\boxed{2}}$}
\put(-5,5.0){$\text{\boxed{1}}$}
}
\caption{Parameter diagram for the cases (1), (2) and (3) in  gradient
  elasticity of bi-Helmholtz type respecting the condition of positive
semidefiniteness of the strain energy density for the material lengths $\ell_1$ and $\ell_2$: $\ell_1^2\ge 0$ and  $\ell_2^4\ge 0$.}
\label{fig:Para}
\end{figure}
An important matter is the question concerning 
the mathematical character of the 
auxiliary lengths $c_1$ and $c_2$.
In Mindlin's theory of second strain gradient elasticity~\cite{Mindlin65,Mindlin72},
Mindlin~\cite{Mindlin65} (see also~\cite{Wu}) pointed out that the conditions for positive
$\mathcal{W}$ supply no indications of the character, real or complex, of the 
four auxiliary lengths  $l_{11}$, $l_{21}$, $l_{12}$, $l_{22}$ of Mindlin's second strain gradient elasticity.
In the theory of gradient elasticity of bi-Helmholtz type,  
the condition for the character, real or complex, of the auxiliary lengths 
$c_1$ and $c_2$ can be obtained from the condition 
if in Eq.~\eqref{c1-2} the discriminant, $1-4\ell_2^4/\ell_1^4$, 
is positive or negative.
We can distinguish three cases according to the sign of the discriminant:
\begin{itemize}
\item[(1)]  
$\ell_1^4>4\ell_2^4$\,:\\
In this case, there are the two auxiliary lengths being real and distinct
\begin{align}
 c_1^2\neq c_2^2\,.
 \end{align}  
This case corresponds to region $\boxed{1}$ in Fig.~\ref{fig:Para}.
The auxiliary lengths $c_1$ and $c_2$ read as 
\begin{align}
\label{c1-2-2-2}
c_{1,2}&=\sqrt{\frac{\ell_1^{2}}{2}\Bigg(1\pm\sqrt{1-4\,\frac{\ell_2^{4}}{\ell_1^{4}}}\Bigg)}\,.
\end{align}
The limit to gradient elasticity of Helmholtz type is: 
$\ell_2^4\rightarrow 0$. 
\item[(2)]
$\ell_1^4=4\ell_2^4$\,: \\
There are the two  auxiliary  lengths being real and equal
\begin{align}
 c_1^2 =c_2^2=\frac{\ell_1^2}{2}\,.
 \end{align}  
This case corresponds to the parabola~$\boxed{2}$  in Fig.~\ref{fig:Para} defined according to $\ell_1^4=4\ell_2^4$.
There is no limit to gradient elasticity of Helmholtz type. 
\item[(3)]
$\ell_1^4<4\ell_2^4$\,:\\
In this region, the two lengths $c_1^2$ and $c_2^2$  are complex conjugate
\begin{align}
\label{c1-2-c2-2}
c^2_{1,2}&=\alpha\pm\ii \beta\,
\end{align}
with 
\begin{align}
\label{alpha}
\alpha=\frac{\ell_1^2}{2}\qquad\text{and}\qquad
\beta=\sqrt{\ell_2^4-\frac{\ell_1^4}{4}}\,.
\end{align}
This case corresponds to region $\boxed{3}$ in Fig.~\ref{fig:Para}.
Moreover, the two auxiliary lengths $c_1$ and $c_2$ are also complex conjugate
and read as
\begin{align}
\label{c1-2-c}
c_{1,2}&=A\pm\ii B\,,
\end{align}
where $A=\text{Re}\, (c_1)=\text{Re}\, (c_2)$
and  $B=\text{Im}\, (c_1)=-\text{Im}\, (c_2)$.
The original material lengths, $\ell_1$ and $\ell_2$ read in terms of $A$ and $B$
\begin{align}
\label{l1-c}
\ell_{1}^2&=2(A^2-B^2)\,,\\
\label{l2-c}
\ell_2^2&=A^2+B^2\,
\end{align}
and the auxiliary lengths $c_1$ and $c_2$ 
\begin{align}
\label{c1-c2}
c^2_{1,2}&=A^2-B^2\pm2\ii AB\,.
\end{align}
Comparing Eq.~\eqref{c1-2-c2-2} with Eq.~\eqref{c1-c2}, one gets
\begin{align}
\label{}
\alpha=A^2-B^2\,,\qquad 
\beta=2 AB\,.
\end{align}
From the first part of the condition of positive semidefiniteness of the strain energy~\eqref{pos-l} and Eq.~\eqref{l1-c}, we obtain
\begin{align}
\label{AB-c}
A^2\ge B^2\,.
\end{align}
Interesting to note that for the particular case $A^2=B^2$, it follows that $\ell_1^2=0$ and the first strain gradients disappear in the strain
energy density~\eqref{W-BH2}. 
The corresponding strain gradient elasticity would be a gradient theory given only in terms of the strain and 
second strain gradient tensors.
The cases $A^2<B^2$ and $A^2=0$ violate the first condition of positive
semidefiniteness of the strain energy density~\eqref{pos-l}.
Moreover Eqs.~\eqref{l1-c} and \eqref{l2-c} give 
 \begin{align}
\label{AB}
A=\ell_2\,\sqrt{\frac{1}{2}+\frac{\ell_1^2}{4\ell_2^2}}\,,\qquad
B=\ell_2\,\sqrt{\frac{1}{2}-\frac{\ell_1^2}{4\ell_2^2}}\,.
\end{align}

For further calculations, it is convenient to introduce the inverse complex
lengths
\begin{align}
\label{c1-2-c-1}
c^{-1}_{1,2}&=a\pm\ii b 
\end{align}
with 
\begin{align}
\label{ab}
a=\frac{A}{A^2+B^2}=\frac{1}{\ell_2}\,\sqrt{\frac{1}{2}+\frac{\ell_1^2}{4\ell_2^2}}\,,\qquad 
b=-\frac{B}{A^2+B^2}=-\frac{1}{\ell_2}\, \sqrt{\frac{1}{2}-\frac{\ell_1^2}{4\ell_2^2}}\,.
\end{align}
Since the auxiliary lengths $c_1$ and $c_2$ are complex conjugate, 
the characteristic material lengths $\ell_1$ and $\ell_2$ are real.
The limit to gradient elasticity of Helmholtz type ($\ell_2^4\rightarrow 0$)
leads to $\ell_1^4<0$,
violating the condition of positive semidefiniteness of the strain 
energy density~\eqref{pos-l}.
\end{itemize}

\subsection{Parameter study}

An important issue in strain gradient elasticity theories is the calculation of 
the characteristic lengths in addition to the elastic constants~\cite{Shodja13,Lazar17b,Po17}.
Zhang et al.~\cite{Zhang} determined, in an atomistic calculation, 
the two auxiliary lengths $c_1$ and $c_2$ as positive and real for graphene.
On the other hand, 
using the Sutton-Chen interatomic potential,
Shodja et al.~\cite{Shodja12} calculated the values of the four auxiliary 
lengths  $l_{11}$, $l_{21}$, $l_{12}$, $l_{22}$ 
of Mindlin's second strain gradient elasticity for fcc crystals 
and found complex conjugate lengths.
Using  ab initio calculations, Ojaghnezhad and Shodja~\cite{Shodja13b} obtained positive and real lengths of Mindlin's second strain gradient elasticity. 
The two auxiliary lengths $c_1$ and $c_2$ of gradient elasticity of bi-Helmholtz type
can be approximated from the four auxiliary lengths
$l_{11}$, $l_{21}$, $l_{12}$, $l_{22}$
of Mindlin's second strain gradient elasticity as
\begin{align}
\label{L-rel1}
c_1&=\frac{l_{11}+l_{21}}{2}\,,\qquad
c_2=\frac{l_{12}+l_{22}}{2}\,.
\end{align}
The two auxiliary lengths $c_1$ and $c_2$ 
might be considered as the average of the four auxiliary lengths $l_{11}$, $l_{21}$
and $l_{12}$, $l_{22}$, respectively.
$c_1^2$, $l^2_{11}$, $l_{21}^2$ correspond to the plus sign in front of the square root
and $c_2^2$, $l^2_{12}$, $l_{22}^2$ correspond to the minus sign in front of the square root
(see Eq.~(\ref{c1-2})  and related expressions in~\cite{Mindlin65,Shodja12}).
Therefore, we study the field solutions for the cases (1), (2) and (3).

Here, we provide the lengths used in this work in the numerical study of three-dimensional gradient elasticity of bi-Helmholtz type:\\
Case (1):\\
\begin{align}
\label{c1-1}
c_{1}=0.4\,a_0\,,
\qquad
c_{2}=0.2\, a_0
\end{align}
and 
\begin{align}
\label{l1-1}
\ell_{1}
=0.447\,a_0\,,
\qquad
\ell_{2}
=0.283\, a_0\,,
\end{align}
where $a_0$ is the lattice constant of the material under consideration.
This choice is motivated by atomistic calculations in first strain gradient elasticity~\cite{Shodja13,Po17,Lazar17b}
and by the value $\ell =0.4\, a_0$ found in the study of dispersion curves in Eringen's nonlocal elasticity~\cite{Eringen83,Eringen}. 
\\
Case (2):\\
\begin{align}
\label{c1-2-2}
c_{1}=c_2=\ell_2=\frac{\ell_1}{\sqrt{2}}=0.283\,a_0\,.
\end{align}
Case (3):\\
\begin{align}
\label{c1-3}
c_{1}=(0.250+ 0.197\,\ii) \,a_0\,,
\qquad
c_{2}=(0.250-0.197\,\ii) \,a_0 
\end{align}
and 
\begin{align}
\label{l1-3}
\ell_{1}
=0.218\,a_0\,,
\qquad
\ell_{2}
=0.318\, a_0\,.
\end{align}
This choice is motivated by atomistic calculations in Mindlin's second strain gradient elasticity~\cite{Shodja12}
and by the study of dispersion curves in nonlocal elasticity of bi-Helmholtz
type~\cite{LMA06b}. Using the values~\eqref{l1-3}, the dispersion curve
in nonlocal elasticity of bi-Helmholtz type coincides with the one based on
the Born-K{\'a}rm{\'a}n model (see~\cite{Kunin,Eringen,LMA06b}).

From the mathematical point of view, the lengths $\ell_1$, $\ell_2$ and $c_1$, $c_2$ play the role of regularization parameters
in gradient elasticity of bi-Helmholtz type.

\subsection{Green tensor of the three-dimensional bi-Helmholtz-Navier equation and relevant Green functions} 

\begin{figure}[t]\unitlength1cm
\vspace*{0.1cm}
\centerline{
\epsfig{figure=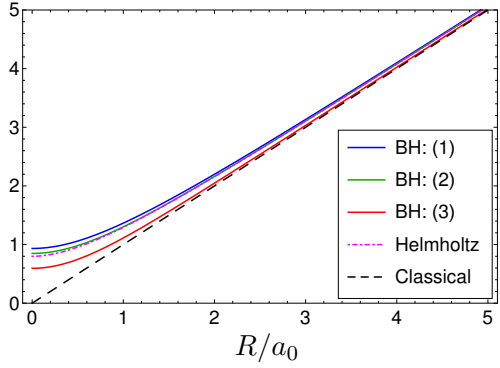,width=6.5cm}
\put(-3.5,-0.4){$R/a_0$}
}
\caption{Plot of the auxiliary function $A(R)$.}
\label{fig:A-r}
\end{figure}

The Green tensor of the three-dimensional bi-Helmholtz-Navier operator, 
$L L_{ik}$, is defined by
\begin{align}
\label{pde-HN}
&L L_{ik} G_{kj}(\Bx-\Bx') =-\delta_{ij} \delta(\Bx-\Bx')
\end{align}
and reads as~\cite{Lazar13} 
\begin{align}
\label{G}
G_{ij}(\bm R)=\frac{1}{16\pi\mu(1-\nu)}\, \Big[2(1-\nu)\delta_{ij}\Delta-
\pd_i\pd_j\Big] A(R)\,
\end{align}
with the auxiliary function 
for the Green tensor of the bi-Helmholtz-Navier operator for case~(1)
\begin{align}
\label{A-BH}
A(R)=R+\frac{2(c_1^2+c_2^2)}{R}
-\frac{2}{c_1^2-c_2^2}\,\frac{1}{R}
\Big(c_1^4\,\e^{-R/c_1}-c_2^4\,\e^{-R/c_2}\Big)\,,
\end{align}
where $R = |\Bx-\Bx'|$ and $\nu$ is the Poisson ratio.
The auxiliary function $A(R)$ may be regarded as a ``regularized  distance
function''. In the far field, it is just $R$ and 
in the near field, it is modified due to gradient parts 
(second and third parts in Eq.~\eqref{A-BH}) (see Fig.~\ref{fig:A-r}).

In case~(2), Eq.~\eqref{A-BH} simplifies to
\begin{align}
\label{A-BH-S}
A(R)=R+\frac{4 c_1^2}{R}
-\bigg(c_1+\frac{4 c_1^2}{R}\bigg)\,\e^{-R/c_1}\,.
\end{align}

In case~(3), using Eq.~\eqref{c1-2-c2-2} and Euler's formula,
Eq.~\eqref{A-BH} reduces to
\begin{align}
\label{A-BH-C}
A(R)=R+\frac{4 (A^2-B^2)}{R}
-\frac{\e^{-a R}}{R} \bigg[4(A^2-B^2)\cos(b R)-\frac{A^4-6A^2B^2+B^4}{AB}\,\sin(b R)
\bigg]\,,
\end{align} 
where $A$ and $B$ are given in Eq.~\eqref{AB}
and $a$ and $b$ are given in Eq.~\eqref{ab}.
The auxiliary function~$A(R)$ is plotted for the cases~(1), (2) and (3) 
in Fig.~\ref{fig:A-r}.

\begin{figure}[t]\unitlength1cm
\vspace*{0.1cm}
\centerline{
\epsfig{figure=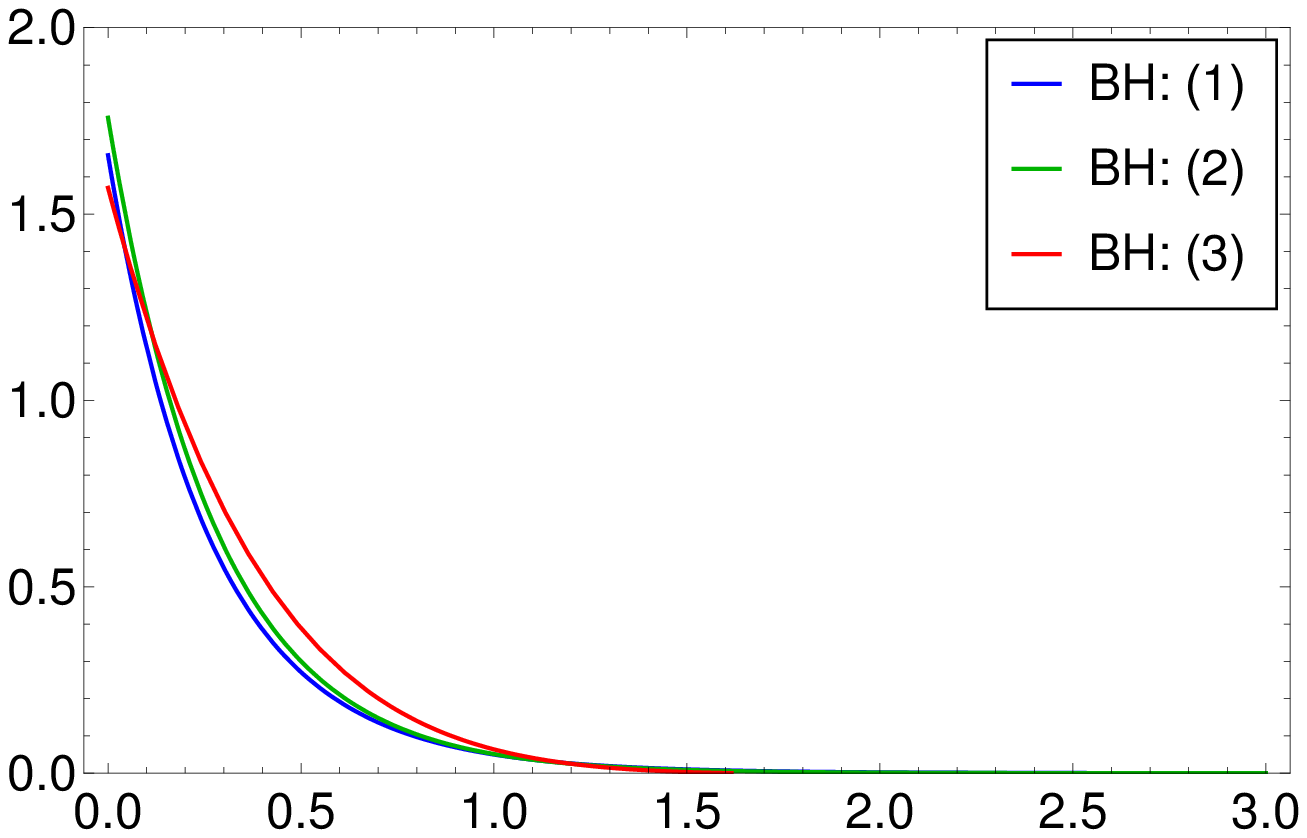,width=6.5cm}
\put(-3.5,-0.4){$R/a_0$}
\put(-6.6,-0.4){$\text{(a)}$}
\hspace*{0.4cm}
\put(0,-0.4){$\text{(b)}$}
\put(3.0,-0.4){$R/a_0$}
\epsfig{figure=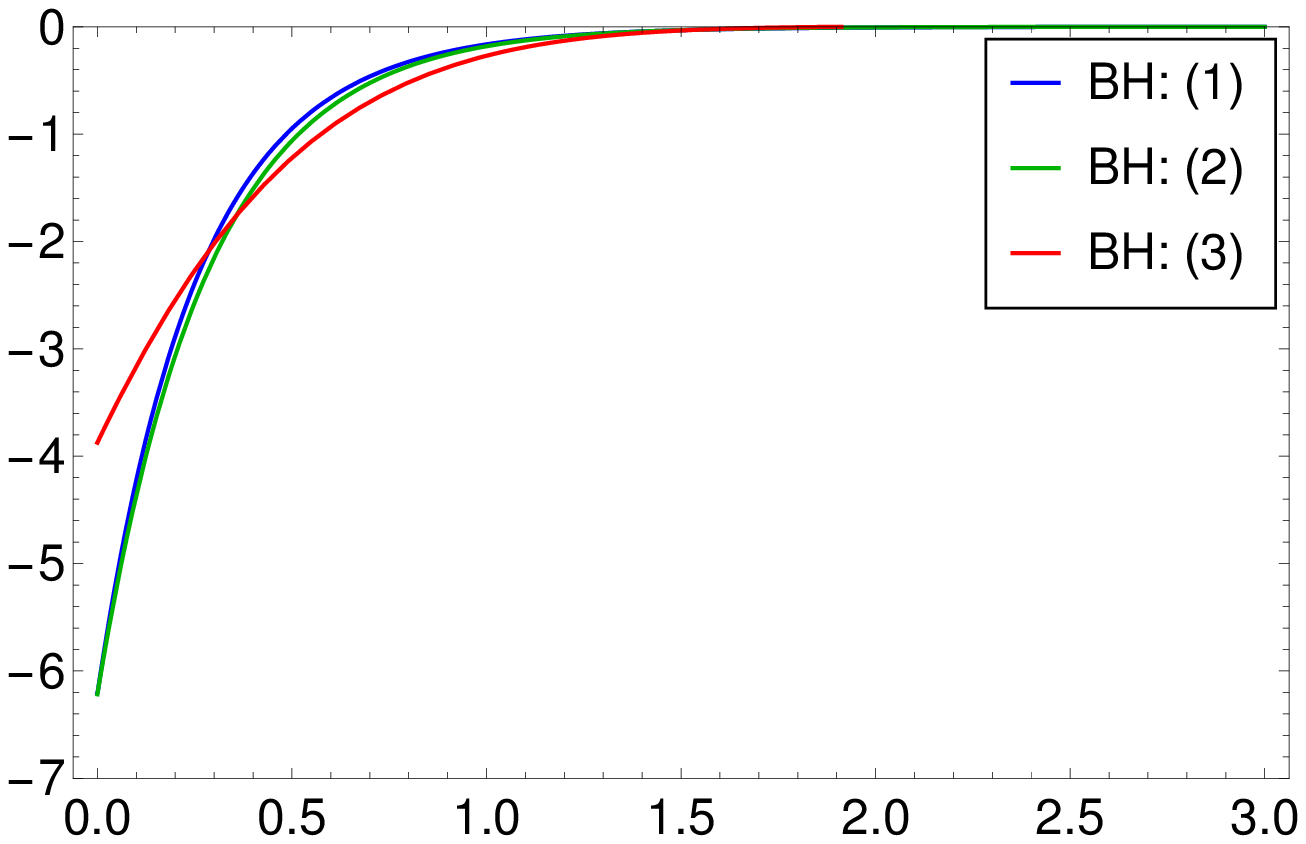,width=6.5cm}
}
\caption{Plot of the Green function: 
(a) $G^{\text{BH}}(R)$ and 
(b) radial part of $\pd_k G^{\text{BH}}(R)$ for the cases (1), (2) and (3).}
\label{fig:GF}
\end{figure}

The Green tensor of the bi-Helmholtz-Navier equation is non-singular 
(see also~\cite{Lazar13}).
Thus, Eq.~(\ref{G}) represents the regularized Green tensor in gradient
elasticity of bi-Helmholtz type.
Note that $A(R)$ can be written as the convolution of $R$ and $G^{\text{BH}}(R)$
\begin{align}
\label{A-reg}
A(R)=R*G^{\text{BH}}(R)\,,
\end{align}
where the symbol $*$ denotes the spatial convolution and 
$G^{\text{BH}}$ is the Green function of the three-dimensional bi-Helmholtz equation
\begin{align}
\label{pde-BH}
L G^{\text{BH}}(R) =\delta(\Bx-\Bx')\,,
\end{align}
which reads for case~(1) (e.g.,~\cite{LMA06})
\begin{align}
\label{G-BH}
G^{\text{BH}}(R)=\frac{1}{4\pi (c_1^2-c_2^2) R}\, 
\Big(\e^{-R/c_1}-\e^{-R/c_2}\Big)\,.
\end{align}
It can be seen in Eq.~(\ref{A-reg}) that $G^{\text{BH}}(R)$ plays the role of 
the regularization function in gradient elasticity of bi-Helmholtz type. 
$G^{\text{BH}}$ is a ``mollifier" (e.g., \cite{Giusti}).
From the physical point of view, $G^{\text{BH}}(R)$ characterizes the shape of the defect and is a
form factor or shape function of the defect. 
The Green function~(\ref{G-BH}) is non-singular and possesses a maximum value at
$R=0$, namely
\begin{align}
\label{G-BH-0}
G^{\text{BH}}(0)=\frac{1}{4\pi c_1 c_2 (c_1+c_2)}\,.
\end{align}
Moreover, the Green function~(\ref{G-BH}) is a Dirac-delta sequence 
with parametric dependence $c_1$ and $c_2$
\begin{align}
\lim_{c_1\to 0,\,  c_2 \to 0} G^{\text{BH}}(R)=\delta(\BR)
\end{align}
and it plays the role of the ``regularization Green function'' 
in gradient elasticity of bi-Helmholtz type. 
The Green function (\ref{G-BH}) is plotted in Fig.~\ref{fig:GF}(a).

In case~(2), the Green function~\eqref{G-BH} simplifies to
\begin{align}
\label{G-BH-2}
G^{\text{BH}}(R)=\frac{1}{8\pi c_1^3}\, \e^{-R/c_1}\,,
\end{align}
possessing a maximum value at $R=0$
\begin{align}
\label{G-BH-0-2}
G^{\text{BH}}(0)=\frac{1}{8\pi c_1^3}\,.
\end{align}

In case~(3), the Green function~\eqref{G-BH} reduces to 
\begin{align}
\label{G-BH-3}
G^{\text{BH}}(R)=-\frac{1}{8\pi AB}\, 
\frac{\e^{-a R}}{R}\, \sin(bR)\,,
\end{align}
having a maximum value at $R=0$
\begin{align}
\label{G-BH-0-3}
G^{\text{BH}}(0)=\frac{1}{8\pi A (A^2+B^2)}\,.
\end{align}
The Green function~$G^{\text{BH}}(R)$ is plotted for the cases~(1), (2) and (3) 
in Fig.~\ref{fig:GF}(a).

The gradient of the Green function~(\ref{G-BH}) of the bi-Helmholtz equation is given by 
\begin{align}
\label{G-grad-BH}
\pd_k G^{\text{BH}}(R)=-\frac{1}{4\pi (c_1^2-c_2^2)}\, 
\frac{R_k}{R^3 }\bigg[\e^{-R/c_1}-\e^{-R/c_2}
+\frac{R}{c_1}\, \e^{-R/c_1}-\frac{R}{c_2}\,\e^{-R/c_2}
\bigg]\,.
\end{align}
Eq.~(\ref{G-grad-BH}) is non-singular 
and possesses a finite value at
$R=0$, namely
\begin{align}
\label{G-grad-BH-0}
\pd_k G^{\text{BH}}(0)=-\frac{1}{8\pi c^2_1 c^2_2}\,\tau_k\,,
\end{align}
where $\tau_k=R_k/R$ are the components of  the unit vector.
The radial part of the gradient of the Green function (\ref{G-grad-BH}) is plotted in Fig.~\ref{fig:GF}(b). 
Moreover, the second gradient of the Green function~(\ref{G-BH}), $\pd_l\pd_k G(R)$,
leads to a singular expression (with $\frac{1}{R}$-singularity).

Moreover, in case~(2) the gradient of the Green function~\eqref{G-grad-BH} simplifies to 
\begin{align}
\label{G-grad-BH-2}
\pd_k G^{\text{BH}}(R)=-\frac{1}{8\pi c_1^4}\, \frac{R_k}{R^3 }\,\e^{-R/c_1}\,.
\end{align}
\\
In case~(3) the gradient of the Green function~\eqref{G-grad-BH} reduces to
\begin{align}
\label{G-grad-BH-3}
\pd_k G^{\text{BH}}(R)=\frac{1}{8\pi}\, \frac{R_k}{R^3 }
\,\e^{-a R}\bigg[\bigg(\frac{1}{AB}+\frac{R}{B(A^2+B^2)}\bigg)\sin(bR)
+\frac{R}{A(A^2+B^2)}\,\cos(bR)\bigg]
\,. 
\end{align}
Eqs.~(\ref{G-grad-BH-2}) and (\ref{G-grad-BH-3}) 
are plotted in Fig.~\ref{fig:GF}(b).

In addition, it holds
\begin{align}
\Delta\Delta  R=-8\pi\, \delta(\Bx-\Bx')\,.
\end{align}
The function~$A(R)$ fulfills the relations
\begin{align}
\label{A-LDD}
L \Delta\Delta A(R)&=-8\pi\, \delta(\Bx-\Bx')\,,\\
\label{A-DD}
\Delta\Delta A(R)&=-8\pi\, G^{\text{BH}}(R)\,,\\
\label{A-LD}
L\Delta  A(R)&=\frac{2}{R}\,,\\
\label{A-D}
\Delta  A(R)&=-8\pi\, G^{L\Delta}(R)\,,\\
\label{A-L}
L  A(R)&=R\,.
\end{align}
Thus, $A(R)$ is the Green function of Eq.~(\ref{A-LDD})
which is a bi-Helmholtz-bi-Laplace equation.
$G^{L\Delta}(R)$ is the Green function of the bi-Helmholtz-Laplace operator
defined by
\begin{align}
\label{pde-LD}
L\Delta  G^{L\Delta}(R) =\delta(\Bx-\Bx')\,
\end{align}
and it reads
\begin{align}
\label{G-LD}
G^{L\Delta}(R)=\frac{1}{4\pi R} \,f_0(R,c_1,c_2)\,
\end{align}
with the auxiliary function for the Green function 
of the bi-Helmholtz-Laplace operator for case~(1)
\begin{align}
\label{f0}
f_0(R,c_1,c_2)=1-\frac{1}{c_1^2-c_2^2} 
\Big[c_1^2\, \e^{-R/c_1}-c_2^2\, \e^{-R/c_2}\Big]\,.
\end{align}
The relevant series expansion (near field) 
of the auxiliary function~\eqref{f0} reads as 
\begin{align}
\label{f0-ser}
f_0(R,c_1,c_2)&=\frac{1}{(c_1+c_2)}\, R-\frac{1}{6 c_1c_2(c_1+c_2)}\, R^3+\mathcal{O}(R^4)\,.
\end{align}
Therefore, 
the auxiliary function $f_0(R,c_1,c_2)$ plays the mathematical role of a multiplicative regularization function in Eq.~\eqref{G-LD}.

In case~(2),  
the auxiliary function~\eqref{f0} becomes 
\begin{align}
\label{f0-2}
f_0(R,c_1,c_1)=1
-\bigg[1+\frac{R}{2c_1}\bigg]\, \e^{-R/c_1}
\end{align} 
with the near field behaviour
\begin{align}
\label{f0-ser-2}
f_0(R,c_1,c_1)&=\frac{1}{2c_1}\, R-\frac{1}{12 c_1^3}\, R^3+\mathcal{O}(R^4)\,.
\end{align}

\begin{figure}[t]\unitlength1cm
\vspace*{0.1cm}
\centerline{
\epsfig{figure=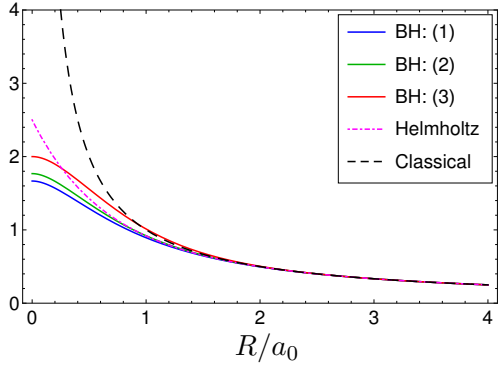,width=6.5cm}
\put(-3.5,-0.4){$R/a_0$}
}
\caption{Plot of the Green function $G^{{L\Delta}}(R)$  for the cases (1), (2) and (3).}
\label{fig:f0-r}
\end{figure}

In case~(3),  
the auxiliary functions~\eqref{f0} reduces to
\begin{align}
\label{f0-3}
f_0(R,c_1,c_2)=1
-\e^{-a R} \bigg[\cos(b R)-\frac{A^2-B^2}{2AB}\,\sin(b R)
\bigg]\, 
\end{align} 
with the near field behaviour
\begin{align}
\label{f0-ser-3}
f_0(R,c_1,c_2)&=\frac{1}{2A}\, R-\frac{1}{12A(A^2+B^2)}\, R^3+\mathcal{O}(R^4)\,.
\end{align}
The Green function~$G^{L\Delta}(R)$ is plotted for the cases~(1), (2) and (3) 
in Fig.~\ref{fig:f0-r}.

The divergence of the Green tensor of the bi-Helmholtz-Navier equation~(\ref{G}) reads
\begin{align}
\label{G-div}
G_{ij,j}(\bm R)=\frac{1-2\nu}{16\pi\mu(1-\nu)}\, \pd_i \Delta\, A(R)\,
\end{align}
and the double-divergence of the Green tensor of the bi-Helmholtz-Navier equation~(\ref{G}) reads\footnote{In classical elasticity,
the double-divergence of the Green tensor of the Navier equation reads
\begin{align*}
G^0_{ij,ij}(\bm R)
=-\frac{1-2\nu}{2\mu(1-\nu)}\, \delta(\bm R)\,,
\end{align*}
which is often erroneously neglected in the literature (e.g., \cite{Bacon,Teodosiu,Balluffi}). }
\begin{align}
\label{G-divdiv}
G_{ij,ij}(\bm R)=\frac{1-2\nu}{16\pi\mu(1-\nu)}\, \Delta \Delta\, A(R)
=-\frac{1-2\nu}{2\mu(1-\nu)}\, G^{\text{BH}}(R)\,.
\end{align}

\subsection{Operator split in gradient elasticity} 

The operator split in gradient elasticity~\cite{Lazar14,RA93} 
is mainly based on the decomposition of a partial differential equation (pde) 
of higher-order into a system of
pdes of lower-order and on the property that the appearing differential 
operator(s) can be written as a product of differential operators of lower-order
(operator-split).
The property that the differential operators commute is also important.
The problem in the theory of defects (for vanishing body forces, $\bm b=0$) in the framework of gradient elasticity
is that both fields, the displacement field $\bm u$ on the left hand side and the plastic distortion $\bm \beta^\P$ as source field 
on the right hand side of the bi-Helmholtz-Navier equation~(\ref{u-L0}) 
are ``a priori'' unknown since they are both non-classical fields (see \cite{Lazar14}). 

The inhomogeneous bi-Helmholtz-Navier equation~(\ref{u-L0}) 
for the displacement field with 
the plastic distortion (or eigendistortion) tensor 
as source field (pde of 6th-order)
\begin{align}
\label{u-L0-RA}
L L_{ik} u_k =C_{ijkl} \pd_j L \beta^{\TP}_{kl}\,
\end{align}
can be decomposed into the following system of partial differential equations,
namely into two uncoupled inhomogeneous bi-Helmholtz equations (pdes of 4th-order)
\begin{align}
\label{u-H-RA}
&L u_i =u_i^0\,,\\
\label{BP-H-RA}
&L \beta^\TP_{ij} =\beta_{ij}^{\TP,0}\,
\end{align}
and into an inhomogeneous Navier equation (pde of 2nd-order)
\begin{align}
\label{u-L0-RA0}
L_{ik} u_k^0 =C_{ijkl} \pd_j  \beta^{\TP,0}_{kl}\,.
\end{align}
Eq.~(\ref{u-L0-RA0}) is the classical Navier equation with the classical
displacement vector $u_k^0$ and the classical eigendistortion tensor 
$\beta^{\TP,0}_{kl}$
known from classical eigenstrain theory,
which are the source fields for Eqs.~(\ref{u-H-RA}) and (\ref{BP-H-RA}).
Substituting Eqs.~(\ref{u-H-RA}) and (\ref{BP-H-RA}) into
Eq.~(\ref{u-L0-RA0}), Eq.~(\ref{u-L0-RA}) is recovered.
Using Eq.~(\ref{BP-H-RA}),
Eq.~(\ref{u-L0-RA}) gives the following bi-Helmholtz-Navier equation
\begin{align}
\label{u-LL}
L L_{ik} u_k &=C_{ijkl} \pd_j  \beta^{\TP,0}_{kl}\,,
\end{align}
where the right hand side is given by the gradient of the classical
eigendistortion tensor. 
Thus, the source term in Eq.~(\ref{u-LL}) is the classical source term known
from Mura's eigenstrain theory (see \cite{Mura}).
Eq.~(\ref{u-LL}) is the basic equation in the eigenstrain formulation 
of gradient elasticity of bi-Helmholtz type.

If we use the Green tensor~(\ref{G}), then the solution 
of Eq.~(\ref{u-LL}) is the convolution of the (negative) Green tensor  with the right hand side of Eq.~(\ref{u-LL}) 
\begin{align}
\label{u-sol}
u_i=G_{ij}*f_j
=-C_{jkln} G_{ij}*\beta_{ln,k}^{\TP,0}
=-C_{jkln} G_{ij,k}*\beta_{ln}^{\TP,0}\,,
\end{align}
where the ficticious body force density reads as 
\begin{align}
\label{f-fic}
f_i=-C_{ijkl} \pd_j  \beta^{\TP,0}_{kl}\,.
\end{align}
Eq.~(\ref{u-sol}) is the generalized Volterra formula for an arbitrary
eigendistortion. 
The gradient of Eq.~(\ref{u-sol}) gives the displacement gradient  
\begin{align}
\label{gradu-sol}
u_{i,m}
=-C_{jkln} G_{ij,km}*\beta_{ln}^{\TP,0}\,.
\end{align}

On the other hand, using the regularization function or mollifier of
bi-Helmholtz type, $G^\text{BH}$, the solutions of Eqs.~\eqref{u-H-RA} and Eqs.~\eqref{BP-H-RA} can be written 
as convolution of the classical singular solution with the mollifier
\begin{align}
\label{u-conv}
u_i&=G^\text{BH}*u_i^0\,,\\
\label{BP-conv}
\beta^\TP_{ij}&=G^\text{BH}*\beta_{ij}^{\TP, 0}\,.
\end{align}

\section{Point defects with cubic symmetry in isotropic materials}
\label{PD}
The aim of the present section is the continuum theoretical modelling of point defects. 
In the theory of incompatible elasticity, point defects can be modelled as
defects corresponding to a three-dimensional Dirac $\delta$-singularity in the
eigendistortion tensor~\cite{Lazar17}.
The eigendistortion or quasi-plastic distortion tensor of a 
point defect with cubic symmetry (dilatation centre) is given by 
\begin{align}
\label{BP-PD-0}
\beta^{\TP,0}_{ij}=Q_{ij}\, \delta(\BR) \,
\quad\text{with}\quad Q_{ij}=Q\,\delta_{ij}\,,
\end{align}
where $\BR=\Bx-\Bx'$ and 
$Q_{}$ is the strength of the cubic point defect (dilatation centre) 
given by
\begin{align}
\label{Q-rel}
Q=\frac{\Delta V}{3}\,.
\end{align}
Here $\Delta V$ represents the plastic volume change~\cite{deWit}.
The point defect, corresponding to the eigendistortion~(\ref{BP-PD-0}), 
is located at point $\Bx'$. 
For point defects~(\ref{BP-PD-0}), 
the  ficticious body force density~(\ref{f-fic}) 
corresponds to double forces (see also \cite{Kossecka71})
\begin{align}
\label{f-fic-PD}
f_i=- P_{ij}\,\pd_j \delta(\BR)\,, \quad
\text{with}\quad
P_{ij}=C_{ijkl} Q_{kl}\,,
\end{align}
where $P_{ij}$ is 
the elastic dipole tensor or double force tensor~\cite{Kroener58,Kroener81,Balluffi,Lazar17}. 

\subsection{Point defects in gradient elasticity of bi-Helmholtz type}
Now, we derive the solutions of the fields of a point defect (dilatation centre) in the framework of 
incompatible gradient elasticity of bi-Helmholtz type.

First, substituting
\begin{align}
C_{ijkk}=2\mu\, \frac{1+\nu}{1-2\nu}\,\delta_{ij}\,
\end{align}
and (\ref{BP-PD-0}) into Eq.~(\ref{u-sol}), we obtain
\begin{align}
\label{u-PDl}
u_i=-\frac{2\mu(1+\nu) Q}{(1-2\nu)} \,G_{ik,k}
=-\frac{Q}{8\pi} \,\frac{1+\nu}{1-\nu}\, \pd_i \Delta\, A(R)\,.
\end{align}
The first gradient of Eq.~(\ref{u-PDl}), being the total distortion tensor, 
reads as
\begin{align}
\label{u-grad-PDl}
u_{i,j}=-\frac{2\mu(1+\nu) Q}{(1-2\nu)} \,G_{ik,kj}
=-\frac{Q}{8\pi} \,\frac{1+\nu}{1-\nu}\, \pd_j \pd_i \Delta\, A(R)\,,
\end{align}
the second gradient of Eq.~(\ref{u-PDl}), 
being the first gradient of the total distortion tensor,
reads as
\begin{align}
\label{u-grad2-PDl}
u_{i,jk}=-\frac{2\mu(1+\nu) Q}{(1-2\nu)} \,G_{im,mjk}
=-\frac{Q}{8\pi} \,\frac{1+\nu}{1-\nu}\,\pd_k \pd_j \pd_i \Delta\, A(R)\,
\end{align}
and the third gradient of Eq.~(\ref{u-PDl}),
being the second gradient of the total distortion tensor,
reads as 
\begin{align}
\label{u-grad3-PDl}
u_{i,jkl}=-\frac{2\mu(1+\nu) Q}{(1-2\nu)} \,G_{im,mjkl}
=-\frac{Q}{8\pi} \,\frac{1+\nu}{1-\nu}\,\pd_l \pd_k \pd_j \pd_i \Delta\, A(R)\,.
\end{align}

\begin{figure}[t]\unitlength1cm
\vspace*{0.1cm}
\centerline{
\epsfig{figure=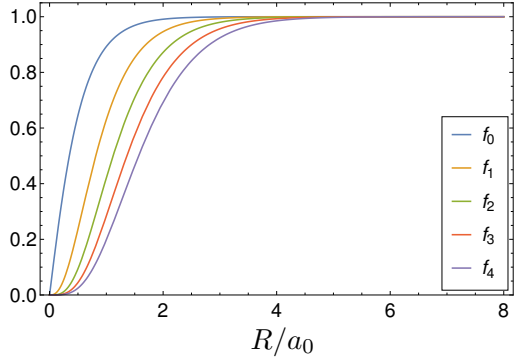,width=6.7cm}
\put(-3.5,-0.4){$R/a_0$}
}
\caption{Plot of the auxiliary functions $f_0$, $f_1$, $f_2$, $f_3$ and $f_4$ for case (1).}
\label{fig:f0-4}
\end{figure}

Introducing the auxiliary functions, playing the mathematical role of multiplicative regularization functions,
\begin{align}
\label{f1}
f_1(R,c_1,c_2)=1
&-\frac{1}{c_1^2-c_2^2}\,\Big[c_1^2\,\e^{-R/c_1}-c_2^2\,\e^{-R/c_2}\Big]
-\frac{R}{c_1^2-c_2^2}\,\Big[c_1\,\e^{-R/c_1}-c_2\,\e^{-R/c_2}\Big]\,,
\\
\label{f2}
f_2(R,c_1,c_2)=1
&-\frac{1}{c_1^2-c_2^2}\,\Big[c_1^2\,\e^{-R/c_1}-c_2^2\,\e^{-R/c_2}\Big]
-\frac{R}{c_1^2-c_2^2}\,\Big[c_1\,\e^{-R/c_1}-c_2\,\e^{-R/c_2}\Big]\nonumber\\
&-\frac{R^2}{3(c_1^2-c_2^2)}\,\Big[\e^{-R/c_1}-\e^{-R/c_2}\Big]\,,
\\
\label{f3}
f_3(R,c_1,c_2)=1
&-\frac{1}{c_1^2-c_2^2}\,\Big[c_1^2\,\e^{-R/c_1}-c_2^2\,\e^{-R/c_2}\Big]
-\frac{R}{c_1^2-c_2^2}\,\Big[c_1\,\e^{-R/c_1}-c_2\,\e^{-R/c_2}\Big]\nonumber\\
&-\frac{2R^2}{5(c_1^2-c_2^2)}\,\Big[\e^{-R/c_1}-\e^{-R/c_2}\Big]
-\frac{R^3}{15(c_1^2-c_2^2)}\,\Big[\frac{1}{c_1}\,
  \e^{-R/c_1}-\frac{1}{c_2}\,\e^{-R/c_2}\Big]\,,
\\
\label{f4}
f_4(R,c_1,c_2)=1
&-\frac{1}{c_1^2-c_2^2}\,\Big[c_1^2\,\e^{-R/c_1}-c_2^2\,\e^{-R/c_2}\Big]
-\frac{R}{c_1^2-c_2^2}\,\Big[c_1\,\e^{-R/c_1}-c_2\,\e^{-R/c_2}\Big]\nonumber\\
&-\frac{3R^2}{7(c_1^2-c_2^2)}\,\Big[\e^{-R/c_1}-\e^{-R/c_2}\Big]
-\frac{2R^3}{21(c_1^2-c_2^2)}\,\Big[\frac{1}{c_1}\, \e^{-R/c_1}-\frac{1}{c_2}\,\e^{-R/c_2}\Big]\nonumber\\
&-\frac{R^4}{105(c_1^2-c_2^2)}\,\Big[\frac{1}{c^2_1}\, \e^{-R/c_1}-\frac{1}{c^2_2}\,\e^{-R/c_2}\Big]\,,
\end{align} 
Eqs.~(\ref{u-PDl})--(\ref{u-grad3-PDl}) can be 
written as
\begin{align}
\label{u-PD}
u_i&=\frac{Q}{4 \pi} \frac{1+\nu}{1-\nu} \frac{R_i}{R^3}\, f_1(R,c_1,c_2)\,,
\\
\label{u-grad-PD}
u_{i,j}&=\frac{Q}{4 \pi} \frac{1+\nu}{1-\nu} \bigg[\frac{\delta_{ij}}{R^3}\,
  f_1(R,c_1,c_2) -\frac{3R_i R_j}{R^5}\, f_2(R,c_1,c_2)\bigg]\,,
\\
\label{u-grad2-PD}
u_{i,jk}&=-\frac{Q}{4 \pi} \frac{1+\nu}{1-\nu} \bigg[\frac{3\big(\delta_{ij}R_k+\delta_{jk}R_i+\delta_{ki}R_j\big)}{R^5}\, f_2(R,c_1,c_2) -\frac{15 R_i R_j R_k}{R^7}\, f_3(R,c_1,c_2)\bigg]\,
\end{align}
and
\begin{align}
\label{u-grad3-PD}
&u_{i,jkl}=-\frac{Q}{4 \pi} \frac{1+\nu}{1-\nu} \bigg[\frac{3\delta_{ijkl}}{R^5}\, f_2(R,c_1,c_2) \nonumber\\
&\quad
-\frac{15\big(\delta_{ij}R_kR_l+\delta_{ik}R_j R_l+\delta_{il}R_jR_k+\delta_{jk}R_iR_l+\delta_{jl}R_iR_k+\delta_{kl}R_iR_j\big)}{R^7}\, f_3(R,c_1,c_2) \nonumber\\
&\quad
+\frac{105 R_i R_j R_k R_l}{R^9}\, f_4(R,c_1,c_2)\bigg]\,
\end{align}
with 
\begin{align}
\label{D-4}
\delta_{ijkl}=\delta_{ij}\delta_{kl}+\delta_{jk}\delta_{il}+\delta_{ik}\delta_{jl}\,.
\end{align}
In addition, the divergence of the displacement field reads as 
\begin{align}
\label{u-div-PD}
u_{i,i}=\frac{Q}{4 \pi} \frac{1+\nu}{1-\nu}\, \frac{3}{R^3}\, \big[f_1(R,c_1,c_2) - f_2(R,c_1,c_2)\big]
=Q\, \frac{1+\nu}{1-\nu} \, G^{\text{BH}}(R)
\end{align}
and its first gradient is 
\begin{align}
\label{u-divgrad-PD}
u_{i,ik}=-\frac{Q}{4 \pi} \frac{1+\nu}{1-\nu}\, \frac{15 R_k}{R^5}\, \big[f_2(R,c_1,c_2) - f_3(R,c_1,c_2)\big]
=Q\, \frac{1+\nu}{1-\nu} \, \pd_k G^{\text{BH}}(R)\,. 
\end{align}
The auxiliary functions~\eqref{f0}, \eqref{f1}--\eqref{f3}
are plotted in Fig.~\ref{fig:f0-4}. 
In the far field, the auxiliary functions~\eqref{f0}, \eqref{f1}--\eqref{f3}
approach 1 and 
in the near field, they are modified due to gradient parts 
and approach 0 at the position $R=0$ (see Fig.~\ref{fig:f0-4}).

Moreover, the quasi-plastic distortion tensor of a dilatation centre reads
\begin{align}
\label{PB-PD}
\beta_{ij}^\TP=G^{\text{BH}}(R)*\beta_{ij}^{\TP,0}
=Q \, \delta_{ij} \, G^{\text{BH}}(R)\,,
\end{align}
the first gradient of quasi-plastic distortion tensor of a dilatation centre is
\begin{align}
\label{PB-grad-PD}
\beta_{ij,k}^\TP=Q \, \delta_{ij} \, \pd_k G^{\text{BH}}(R)\,
\end{align}
and the second gradient of quasi-plastic distortion tensor of a dilatation centre reads
\begin{align}
\label{PB-grad2-PD}
\beta_{ij,kl}^\TP=Q \, \delta_{ij} \, \pd_l \pd_k G^{\text{BH}}(R)\,.
\end{align}
In Eq.~(\ref{PB-PD}), it can be seen that  the form factor $G^{\text{BH}}$  characterizes the shape 
of the quasi-plastic distortion of a dilatation centre.

For a dilatation centre, 
the elastic strain tensor and its gradients read
\begin{align}
\label{strain}
e_{ij}&=u_{i,j}-\beta_{ij}^\TP\,,
\\
\label{strain-grad}
e_{ij,k}&=u_{i,jk}-\beta_{ij,k}^\TP\,,
\\
\label{strain-grad2}
e_{ij,kl}&=u_{i,jkl}-\beta_{ij,kl}^\TP\,
\end{align}
and the elastic dilatation field and its gradients are 
\begin{align}
\label{dil}
e_{ii}&=-Q\, \frac{2(1-2\nu)}{1-\nu} \, G^{\text{BH}}(R)\,,
\\
\label{dil-grad}
e_{ii,k}&=-Q\, \frac{2(1-2\nu)}{1-\nu} \, \pd_k G^{\text{BH}}(R)\,,
\\
\label{dil-grad2}
e_{ii,kl}&=-Q\, \frac{2(1-2\nu)}{1-\nu} \, \pd_l \pd_k G^{\text{BH}}(R)\,.
\end{align}

In order to study if the fields are singularity-free,
the near-field behaviour of the fields~(\ref{u-PD})--(\ref{u-grad3-PD})
is needed. 
The relevant series expansion of the auxiliary functions~\eqref{f1}--\eqref{f4}
(near fields) reads as
\begin{align}
\label{f1-ser}
f_1(R,c_1,c_2)&=\frac{1}{3 c_1c_2 (c_1+c_2)}\, R^3-\frac{1}{8 c^2_1c^2_2}\, R^4+\mathcal{O}(R^5)\,,\\
\label{f2-ser}
f_2(R,c_1,c_2)&=\frac{1}{24 c^2_1c^2_2}\, R^4 +\mathcal{O}(R^5)\,,\\
\label{f3-ser}
f_3(R,c_1,c_2)&=\frac{1}{120 c^2_1c^2_2}\, R^4 +\mathcal{O}(R^5)\,,\\
\label{f4-ser}
f_4(R,c_1,c_2)&=\frac{1}{280 c^2_1c^2_2}\, R^4 +\mathcal{O}(R^5)\,.
\end{align}
From Eqs.~\eqref{f1-ser}--\eqref{f4-ser}
it can be seen that the function $f_1(R,c_1,c_2)$ regularizes up to a
$1/R^3$-singularity and the functions $f_2(R,c_1,c_2)$,  $f_3(R,c_1,c_2)$
and  $f_4(R,c_1,c_2)$ regularize up to a
$1/R^4$-singularity and give non-singular fields.

\begin{figure}[t]\unitlength1cm
\vspace*{0.1cm}
\centerline{
\epsfig{figure=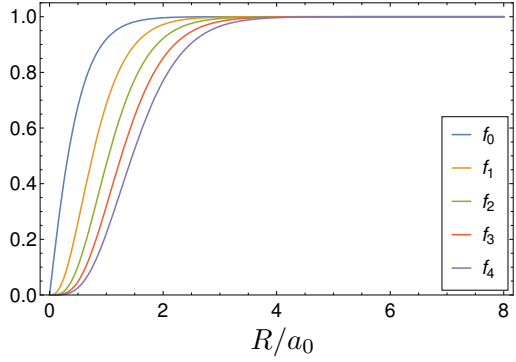,width=6.7cm}
\put(-3.5,-0.4){$R/a_0$}
}
\caption{Plot of the auxiliary functions $f_0$, $f_1$, $f_2$, $f_3$ and $f_4$ for case (2).}
\label{fig:f0-4-2}
\end{figure}

In case~(2),  
the auxiliary functions~\eqref{f1}--\eqref{f4} become 
\begin{align}
\label{f1-2}
f_1(R,c_1,c_1)=1
&-\bigg[1+\frac{R}{c_1}+\frac{R^2}{2c_1^2}\bigg]\, \e^{-R/c_1}\,,
\\
\label{f2-2}
f_2(R,c_1,c_1)=1
&-\bigg[1+\frac{R}{c_1}+\frac{R^2}{2c_1^2}+\frac{R^3}{6c_1^3}\bigg]\,
\e^{-R/c_1}\,,
\\
\label{f3-2}
f_3(R,c_1,c_1)=1
&-\bigg[1+\frac{R}{c_1}+\frac{R^2}{2c_1^2}+\frac{R^3}{6c_1^3}+\frac{R^4}{30c_1^4}\bigg]\, \e^{-R/c_1}\,,
\\
\label{f4-2}
f_4(R,c_1,c_1)=1
&-\bigg[1+\frac{R}{c_1}+\frac{R^2}{2c_1^2}+\frac{R^3}{6c_1^3}
+\frac{4R^4}{105c_1^4}+\frac{R^5}{210c_1^5}\bigg]\, \e^{-R/c_1}\,,
\end{align} 
where the relevant series expansion of the auxiliary 
functions~\eqref{f1-2}--\eqref{f4-2} (near fields) reads as 
\begin{align}
\label{f1-ser-2}
f_1(R,c_1,c_1)&=\frac{1}{6 c^3_1}\, R^3-\frac{1}{8 c^4_1}\, R^4+\mathcal{O}(R^5)\,,\\
\label{f2-ser-2}
f_2(R,c_1,c_1)&=\frac{1}{24 c^4_1}\, R^4 +\mathcal{O}(R^5)\,,\\
\label{f3-ser-2}
f_3(R,c_1,c_1)&=\frac{1}{120 c^4_1}\, R^4 +\mathcal{O}(R^5)\,,\\
\label{f4-ser-2}
f_4(R,c_1,c_1)&=\frac{1}{280 c^4_1}\, R^4 +\mathcal{O}(R^5)\,.
\end{align}
From Eqs.~\eqref{f1-ser-2}--\eqref{f4-ser-2}
it is obvious that the function $f_1(R,c_1,c_1)$ regularizes up to a
$1/R^3$-singularity and the functions $f_2(R,c_1,c_1)$,  $f_3(R,c_1,c_1)$
and  $f_4(R,c_1,c_1)$ regularize up to a $1/R^4$-singularity. 
The auxiliary functions~\eqref{f0-2}, \eqref{f1-2}--\eqref{f3-2}
are plotted in Fig.~\ref{fig:f0-4-2}.

\begin{figure}[t]\unitlength1cm
\vspace*{0.1cm}
\centerline{
\epsfig{figure=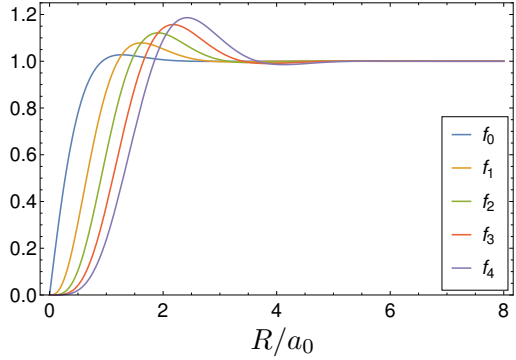,width=6.7cm}
\put(-3.5,-0.4){$R/a_0$}
}
\caption{Plot of the auxiliary functions $f_0$, $f_1$, $f_2$, $f_3$ and $f_4$ for case (3).}
\label{fig:f0-4-3}
\end{figure}

In case~(3),  
the auxiliary functions~\eqref{f1}--\eqref{f4} reduce to 
\begin{align}
\label{f1-3}
f_1(R,c_1,c_2)=1
&-\e^{-a R} \bigg[\bigg(1+\frac{R}{2A}\bigg)\cos(b R)
-\bigg(\frac{A^2-B^2}{2AB}+\frac{R}{2B}\bigg)\sin(b R)
\bigg]\,,
\\
\label{f2-3}
f_2(R,c_1,c_2)=1
&-\e^{-a R} \bigg[\bigg(1+\frac{R}{2A}\bigg)\cos(b R)
-\bigg(\frac{A^2-B^2}{2AB}+\frac{R}{2B}+\frac{R^2}{6AB}\bigg)\sin(b R)
\bigg]\,,
\\
\label{f3-3}
f_3(R,c_1,c_2)=1
&-\e^{-a R} \bigg[\bigg(1+\frac{R}{2A}-\frac{R^3}{30A(A^2+B^2)}\bigg)\cos(b R)
\nonumber\\
&\qquad 
-\bigg(\frac{A^2-B^2}{2AB}+\frac{R}{2B}+\frac{R^2}{5AB}
+\frac{R^3}{30B(A^2+B^2)}\bigg)\sin(b R)
\bigg]\,,
\\
\label{f4-3}
f_4(R,c_1,c_2)=1
&-\e^{-a R} \bigg[\bigg(1+\frac{R}{2A}-\frac{R^3}{21A(A^2+B^2)}
-\frac{R^4}{105(A^2+B^2)^2}\bigg)\cos(b R)
\nonumber\\
&
\hspace*{-2cm}
-\bigg(\frac{A^2-B^2}{2AB}+\frac{R}{2B}+\frac{3R^2}{14AB}
+\frac{R^3}{21B(A^2+B^2)}
+\frac{R^4(A^2-B^2)}{210AB(A^2+B^2)^2}
\bigg)\sin(b R)
\bigg]
\end{align} 
and the corresponding series expansion of the auxiliary 
functions~\eqref{f1-3}--\eqref{f4-3} (near fields) reads as 
\begin{align}
\label{f1-ser-3}
f_1(R,c_1,c_2)&=\frac{1}{6 A(A^2+B^2)}\, R^3-\frac{1}{8 (A^2+B^2)^2}\, R^4+\mathcal{O}(R^5)\,,\\
\label{f2-ser-3}
f_2(R,c_1,c_2)&=\frac{1}{24 (A^2+B^2)^2}\, R^4 +\mathcal{O}(R^5)\,,\\
\label{f3-ser-3}
f_3(R,c_1,c_2)&=\frac{1}{120 (A^2+B^2)^2}\, R^4 +\mathcal{O}(R^6)\,,\\
\label{f4-ser-3}
f_4(R,c_1,c_2)&=\frac{1}{280 (A^2+B^2)^2}\, R^4 +\mathcal{O}(R^5)\,,
\end{align}
where $A$ and $B$ are given in Eq.~\eqref{AB}
and $a$ and $b$ are given in Eq.~\eqref{ab}.
From Eqs.~\eqref{f1-ser-3}--\eqref{f4-ser-3}
it can be seen that the function $f_1(R,c_1,c_2)$ regularizes up to a
$1/R^3$-singularity and the functions $f_2(R,c_1,c_2)$,  $f_3(R,c_1,c_2)$
and  $f_4(R,c_1,c_2)$ regularize up to a $1/R^4$-singularity
and give non-singular fields. 
The auxiliary functions~\eqref{f0-3}, \eqref{f1-3}--\eqref{f3-3}
are plotted in Fig.~\ref{fig:f0-4-3}. 
It can be seen that
the auxiliary functions ~\eqref{f0-3}, \eqref{f1-3}--\eqref{f3-3}
show a weak oscillation in the near field.

\begin{figure}[p]\unitlength1cm
\vspace*{-0.5cm}
\centerline{
\epsfig{figure=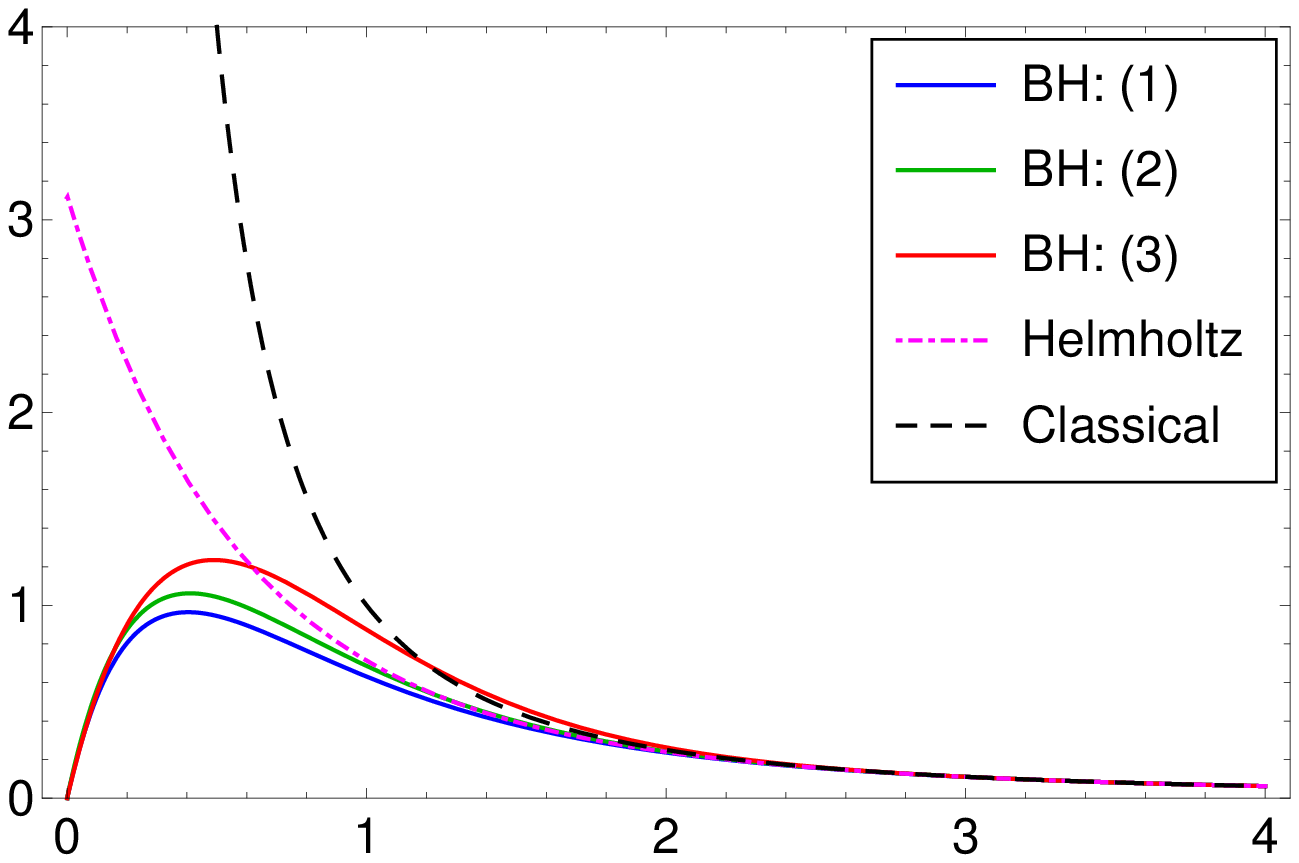,width=6.4cm}
\put(-3.5,-0.4){$R/a_0$}
\put(-6.8,-0.3){$\text{(a)}$}
\hspace*{0.2cm}
\put(-0.1,-0.3){$\text{(b)}$}
\put(3.0,-0.4){$R/a_0$}
\epsfig{figure=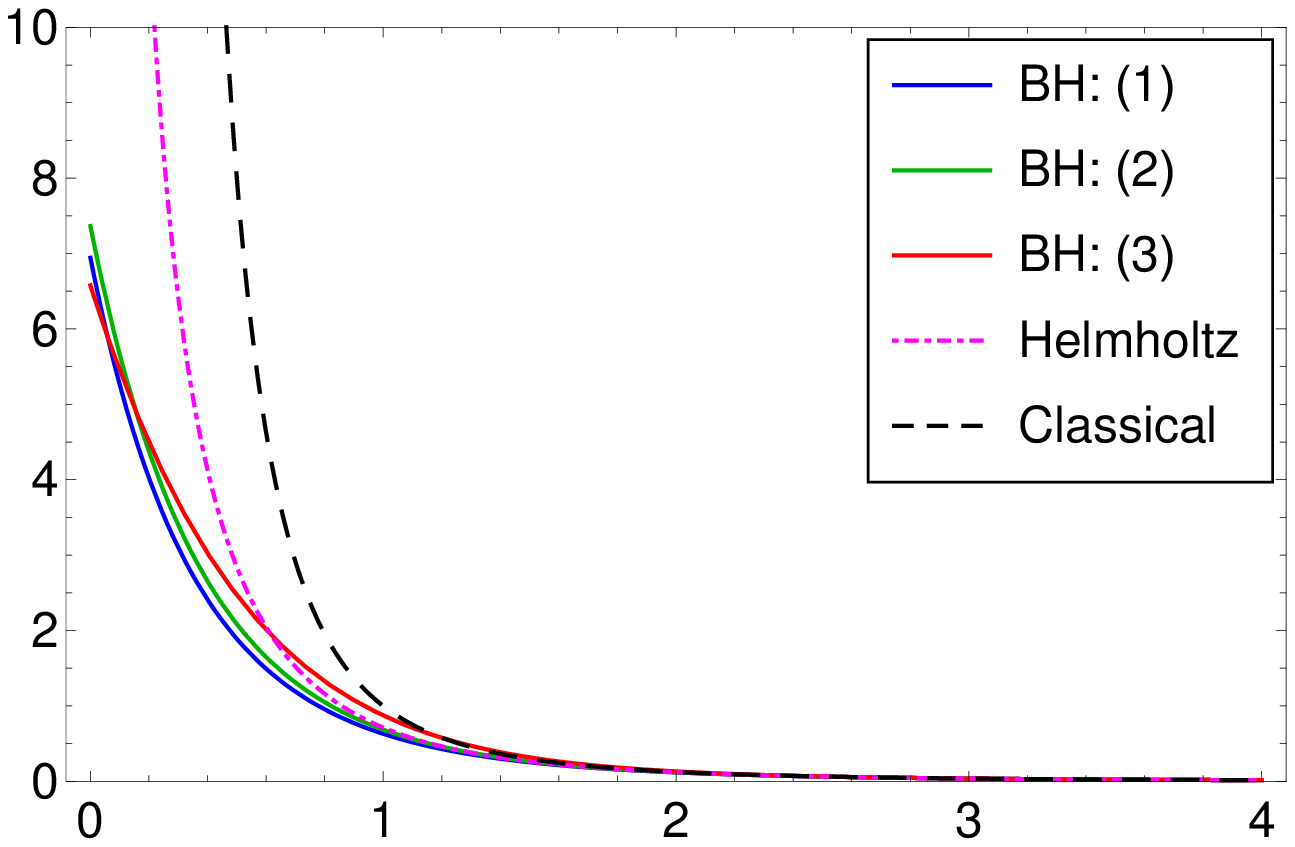,width=6.5cm}
}
\centerline{
\epsfig{figure=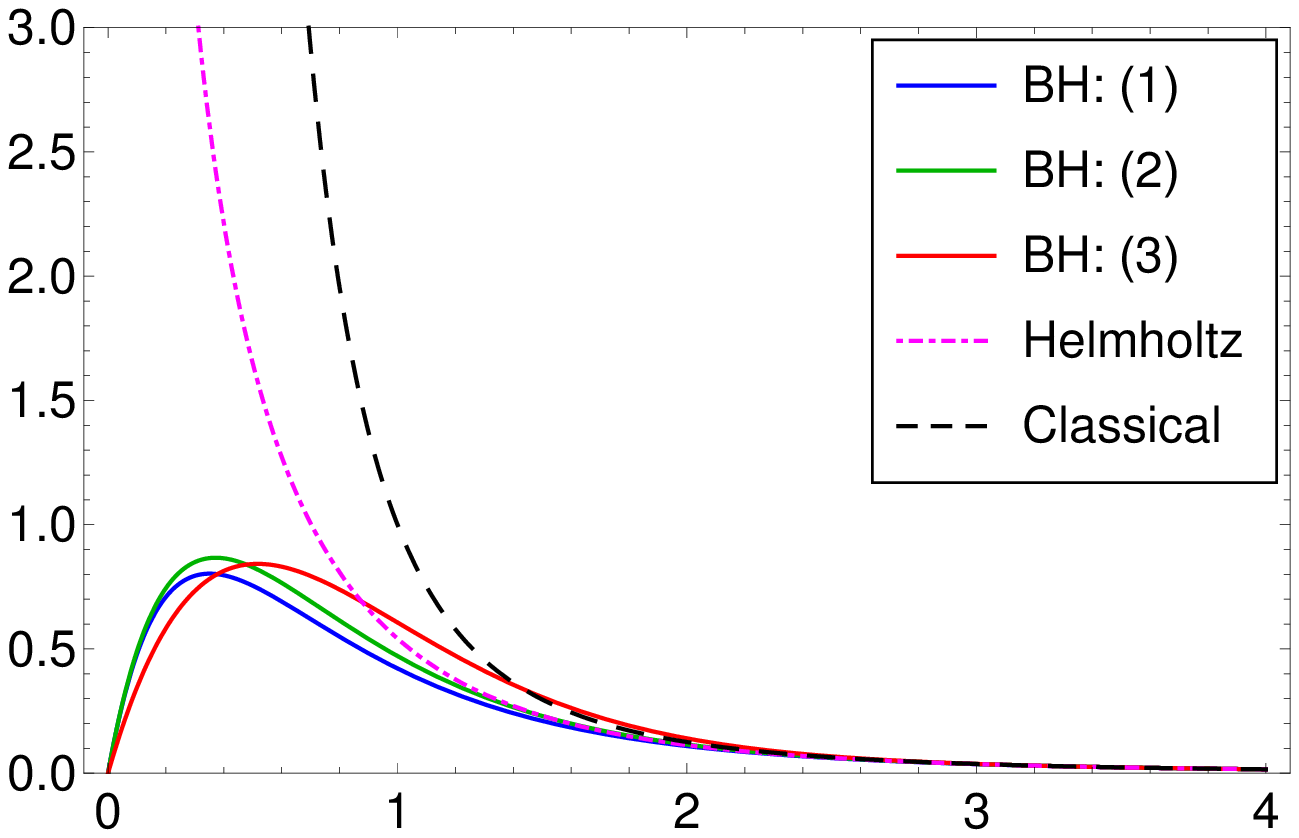,width=6.5cm}
\put(-3.5,-0.4){$R/a_0$}
\put(-6.8,-0.3){$\text{(c)}$}
\hspace*{0.2cm}
\put(3.0,-0.4){$R/a_0$}
\put(-0.1,-0.3){$\text{(d)}$}
\epsfig{figure=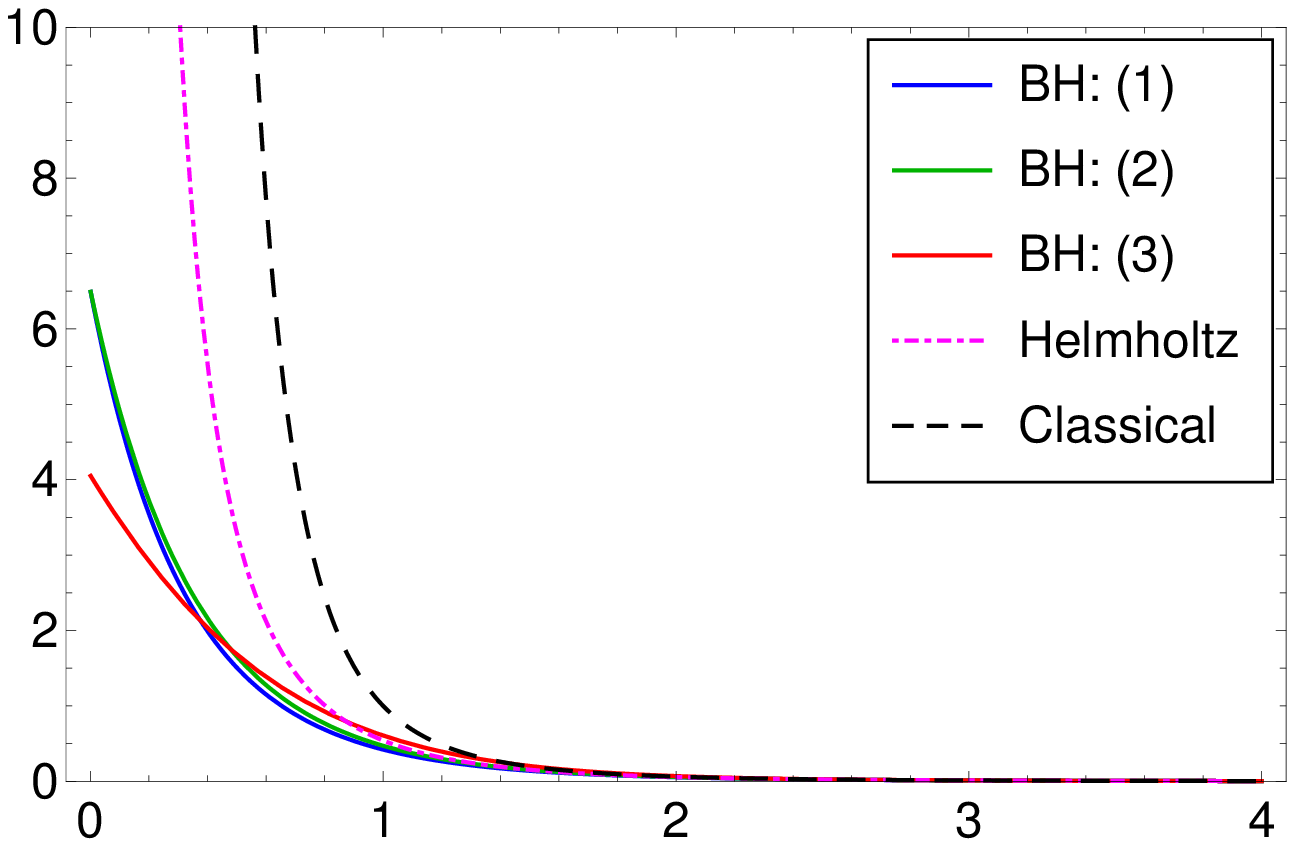,width=6.5cm}
}
\centerline{
\epsfig{figure=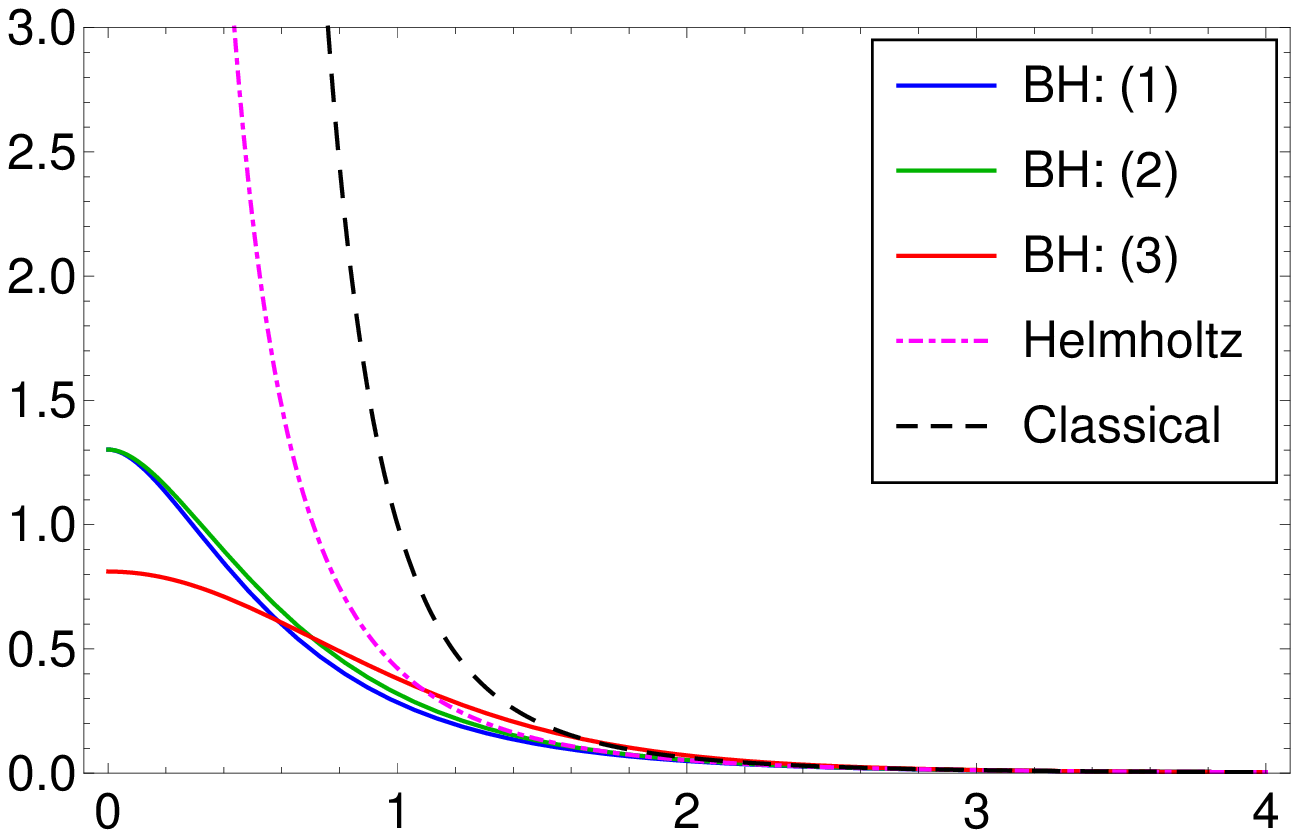,width=6.5cm}
\put(-3.5,-0.4){$R/a_0$}
\put(-6.8,-0.3){$\text{(e)}$}
\hspace*{0.2cm}
\put(-0.1,-0.3){$\text{(f)}$}
\put(3.0,-0.4){$R/a_0$}
\epsfig{figure=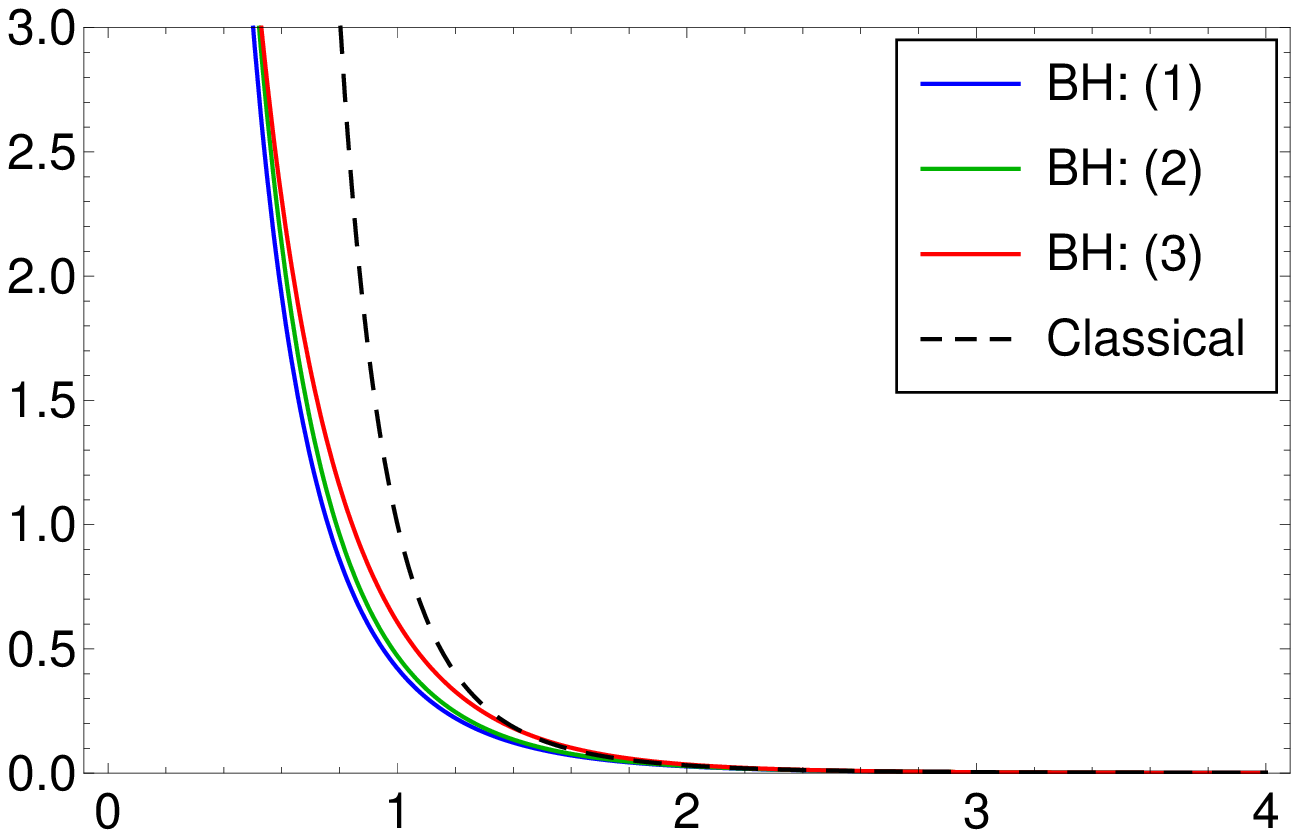,width=6.5cm}
}
\centerline{
\epsfig{figure=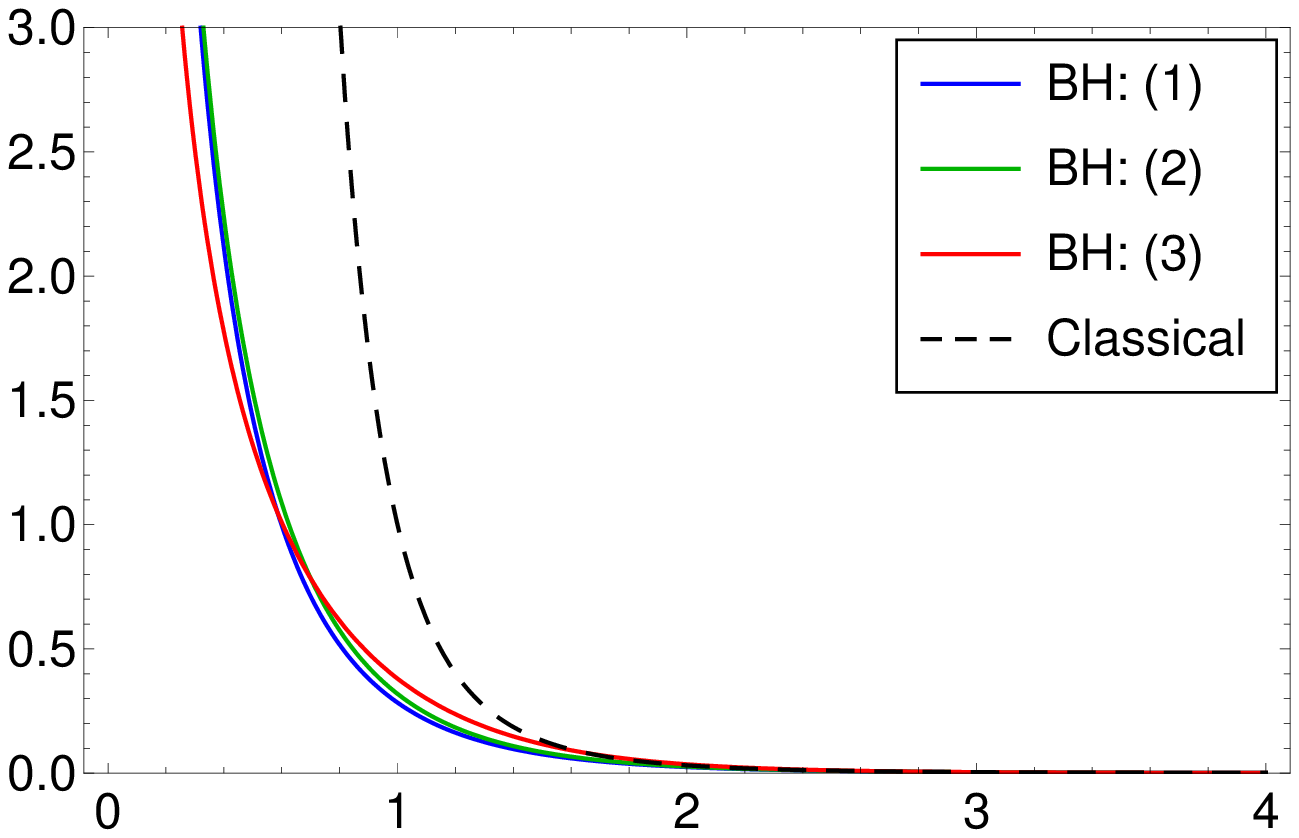,width=6.5cm}
\put(-3.5,-0.4){$R/a_0$}
\put(-6.8,-0.3){$\text{(g)}$}
\hspace*{0.2cm}
\put(3.0,-0.4){$R/a_0$}
\put(-0.1,-0.3){$\text{(h)}$}
\epsfig{figure=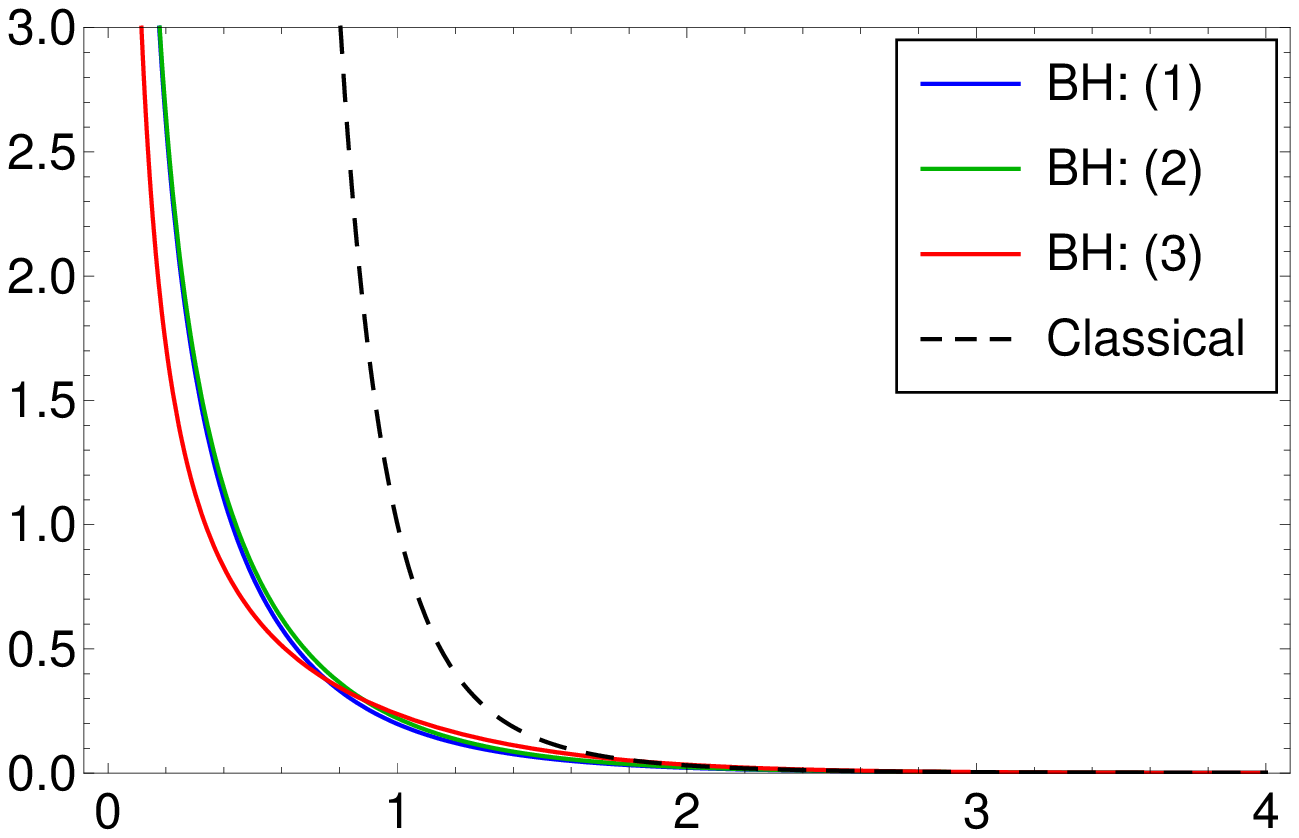,width=6.5cm}
}
\caption{Plots of the radial functions
in gradient elasticity of bi-Helmholtz type,
gradient elasticity of Helmholtz type
and classical elasticity:
(a) $f_1/R^2$,
(b) $f_1/R^3$,
(c) $f_2/R^3$,
(d) $f_2/R^4$,
(e) $f_3/R^4$,
(f) $f_2/R^5$,
(g) $f_3/R^5$,
(h) $f_4/R^5$.}
\label{fig:f}
\end{figure}

Therefore, up to the second gradient of the displacement, 
the fields are non-singular. Only, the third gradient of the displacement
field possesses a $1/R$-singularity. 
The radial part $f_1(R,c_1,c_2)/R^2$ of the displacement (\ref{u-PD}) is zero
at $R=0$, since the near field behaves like $R$, and is plotted 
in Fig.~\ref{fig:f}(a). 
The radial parts $f_1(R,c_1,c_2)/R^3$ and  $f_2(R,c_1,c_2)/R^3$
of the first displacement gradient~(\ref{u-grad-PD}) are finite and 
zero, respectively, at $R=0$ and are plotted in Figs.~\ref{fig:f}(b) and \ref{fig:f}(c). 
The radial parts $f_2(R,c_1,c_2)/R^4$ and  $f_3(R,c_1,c_2)/R^4$
of the second displacement gradient~(\ref{u-grad2-PD}) are finite 
at $R=0$ and are plotted in Figs.~\ref{fig:f}(d) and \ref{fig:f}(e). 
That means that the term
$(\delta_{ij}\tau_k+\delta_{jk}\tau_i+\delta_{ki}\tau_j)f_2(R,c_1,c_2)/R^4$
of the second displacement gradient~(\ref{u-grad2-PD}) has a jump at $R=0$. 
The radial parts $f_2(R,c_1,c_2)/R^5$, $f_3(R,c_1,c_2)/R^5$ and $f_4(R,c_1,c_2)/R^5$
of the third displacement gradient~(\ref{u-grad3-PD}) possess a 
$1/R$-singularity at $R=0$ 
and are plotted in Figs.~\ref{fig:f}(f), \ref{fig:f}(g) 
and \ref{fig:f}(h).
The radial functions and the corresponding ``classical'' singularities
$1/R^2$, $1/R^3$, $1/R^4$ and $1/R^5$ are plotted in Fig.~\ref{fig:f}.
In Fig.~\ref{fig:f}, it can be seen that the functions
$f_1(R,c_1,c_2)/R^2$, 
$f_1(R,c_1,c_2)/R^3$,
$f_2(R,c_1,c_2)/R^3$,
$f_2(R,c_1,c_2)/R^4$,
$f_3(R,c_1,c_2)/R^4$,
$f_2(R,c_1,c_2)/R^5$, 
$f_3(R,c_1,c_2)/R^5$ 
and $f_4(R,c_1,c_2)/R^5$
do not show weak oscillations in the near field.

Using the 3D Green function of the bi-Helmholtz equation, $G^\text{BH}(R)$,
as regularization function with 
$G^\text{BH}(R)*1/R^n$,
gradient elasticity of bi-Helmholtz type regularizes classical singularities up to 
$1/R^4$, namely  $G^\text{BH}(R)*1/R^4$ is finite. 
For the order $n=5$, the classical singularity $1/R^5$ is regularized to a weaker singularity, 
namely $G^\text{BH}(R)*1/R^5\sim 1/R$.

Moreover, substituting Eqs.~\eqref{u-grad-PD}, \eqref{PB-PD} and \eqref{dil} into Eq.~\eqref{CR1}, and using Eq.~\eqref{C} and $\lambda=2\mu\nu/(1-2\nu)$,
the Cauchy stress of a dilatation centre in gradient elasticity of bi-Helmholtz type  is obtained as 
\begin{align} 
\label{sigma-PD-BH}
\sigma_{ij}&=\frac{\mu Q}{2 \pi} \,\frac{1+\nu}{1-\nu} \bigg[
\bigg(\frac{\delta_{ij}}{R^3}-\frac{3R_i R_j}{R^5}\bigg) f_2(R,c_1,c_2)
  -\frac{8\pi}{3}\, \delta_{ij}\, G^\text{BH}(R) \bigg]\,,
\end{align}
which is non-singular. In addition, Eq.~\eqref{sigma-PD-BH} gives also the ``nonlocal" stress of a dilatation centre in 
the framework of nonlocal elasticity of bi-Helmholtz type~\cite{LMA06b}.

\subsection{Limit to gradient elasticity of Helmholtz type}

In the limit from gradient elasticity of bi-Helmholtz type (case (1))
to gradient elasticity of Helmholtz type (e.g.~\cite{AA97,LM05}),
$c_2\rightarrow 0$ and $c_1\rightarrow\ell$,
we obtain from Eqs.~(\ref{u-PDl})--(\ref{u-grad2-PDl}) or Eqs.~(\ref{u-PD})--(\ref{u-divgrad-PD}) the 
displacement field of a dilatation centre in the framework of 
gradient elasticity of Helmholtz type
\begin{align}
\label{u-PD-H}
u_i=-\frac{Q}{4 \pi} \frac{1+\nu}{1-\nu}\, \pd_i \bigg(\frac{1}{R}- \frac{\e^{-R/\ell}}{R}\bigg)
=\frac{Q}{4 \pi} \frac{1+\nu}{1-\nu} \frac{R_i}{R^3}\, f_1(R,\ell)\,,
\end{align}
the first gradient 
\begin{align}
\label{u-grad-PD-H}
u_{i,j}&=-\frac{Q}{4 \pi} \frac{1+\nu}{1-\nu}\, \pd_j\pd_i \bigg(\frac{1}{R}- \frac{\e^{-R/\ell}}{R}\bigg)\nonumber\\
&=\frac{Q}{4 \pi} \frac{1+\nu}{1-\nu} \bigg[\frac{\delta_{ij}}{R^3}\, f_1(R,\ell) -\frac{3R_i R_j}{R^5}\, f_2(R,\ell)\bigg]\,,
\end{align}
which can be verified by making use of the identities~\cite{Rogula73,Frahm}
\begin{align}
\label{diff-1} 
\pd_j\pd_i\bigg(\frac{1}{R}\bigg)&=
-\frac{4\pi}{3}\,\delta_{ij}\,\delta(\BR)
-\frac{\delta_{ij}}{R^3}
+\frac{3R_i R_j}{R^5}\,,\\
\label{diff-2}
\pd_j\pd_i 
\bigg(\frac{\text{e}^{-R/\ell}}{R}\bigg)
&=-\frac{4\pi}{3}\, \delta_{ij}\,\delta(\bm R)
-\frac{\delta_{ij}}{R^3}\Big(1+\frac{R}{\ell}\Big)\text{e}^{-R/\ell}
+3\frac{R_{i} R_j}{R^5}\Big(1+\frac{R}{\ell}+\frac{R^2}{3\ell^2}\Big)\text{e}^{-R/\ell}\,,
\end{align}
the second gradient 
\begin{align}
\label{u-grad2-PD-H}
u_{i,jk}=-\frac{Q}{4 \pi} \frac{1+\nu}{1-\nu} \bigg[\frac{3\big(\delta_{ij}R_k+\delta_{jk}R_i+\delta_{ki}R_j\big)}{R^5}\, f_2(R,\ell) -\frac{15 R_i R_j R_k}{R^7}\, f_3(R,\ell)\bigg]\,
\end{align}
and the divergence of the displacement field reads
\begin{align}
\label{u-div-PD-H}
u_{i,i}=\frac{Q}{4 \pi} \frac{1+\nu}{1-\nu}\, \frac{3}{R^3}\, \big[f_1(R,\ell) - f_2(R,\ell)\big]
=Q\, \frac{1+\nu}{1-\nu} \, G^\text{H}(R)\,,
\end{align}
with the Green function of the Helmholtz operator given by
\begin{align}
\label{G-H}
G^\text{H}(R)=\frac{1}{4\pi\ell^2 R}\,  {\e^{-R/\ell}}\,,
\end{align}
which is a Dirac-delta sequence 
with parametric dependence $\ell$
\begin{align}
\label{G-H-limit}
\lim_{\ell \to 0} G^{\text{H}}(R)=\delta(\BR)\,.
\end{align}
Eq.~\eqref{G-H} is the regularization function in gradient elasticity of Helmholtz type and is a ``mollifier".
The quasi-plastic distortion of a dilatation centre (\ref{PB-PD})
reduces to
\begin{align}
\label{BP-H}
\beta^\TP_{ij}=Q\, \delta_{ij}\,G^\text{H}(R)=\frac{Q}{4\pi\ell^2 R}\, \delta_{ij}\, \e^{-R/\ell}\,.
\end{align}
For gradient elasticity of Helmholtz type, 
the Green function of the Helmholtz-Laplace operator 
is
\begin{align}
\label{G-LD-H}
G^{L\Delta}(R)=\frac{1}{4\pi R} \,f_0(R,\ell)\,
\end{align}
obtained from Eqs.~\eqref{G-LD} and \eqref{f0}.

\begin{figure}[t]\unitlength1cm
\vspace*{0.1cm}
\centerline{
\epsfig{figure=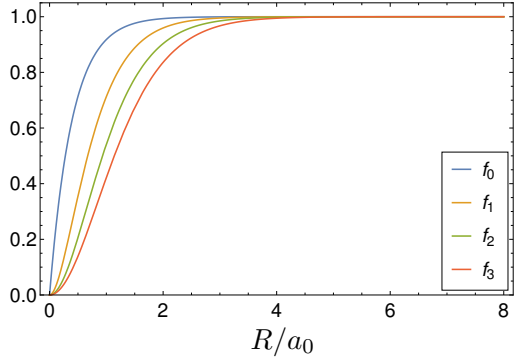,width=6.7cm}
\put(-3.5,-0.4){$R/a_0$}
}
\caption{Plot of the auxiliary functions $f_0$, $f_1$, $f_2$ and $f_3$ in gradient elasticity of Helmholtz type.}
\label{fig:f0-4-H}
\end{figure}

In the limit $c_2\rightarrow 0$ and $c_1\rightarrow\ell$, 
the auxiliary functions~\eqref{f0}, (\ref{f1})--(\ref{f3}) read as
\begin{align}
\label{f0-H}
f_0(R,\ell)&=1-\e^{-R/\ell}\,,\\
\label{f1-H}
f_1(R,\ell)&=1-\bigg[1+\frac{R}{\ell}\bigg]\,\e^{-R/\ell}\,,
\\
\label{f2-H}
f_2(R,\ell)&=1-
\bigg[1+\frac{R}{\ell}+\frac{1}{3}\,\frac{R^2}{\ell^2}\bigg]\,\e^{-R/\ell}\,,
\\
\label{f3-H}
f_3(R,\ell)&=1-
\bigg[1+\frac{R}{\ell}+\frac{2}{5}\,\frac{R^2}{\ell^2}+\frac{1}{15}\,\frac{R^3}{\ell^3}\bigg]\,\e^{-R/\ell}\,.
\end{align} 
The auxiliary functions~(\ref{f0-H})--(\ref{f3-H})
are plotted in Fig.~\ref{fig:f0-4-H} for $\ell=0.4 a_0$. 
For gradient elasticity of Helmholtz type,
the near field behaviour of the fields (\ref{u-PD-H})--(\ref{u-grad2-PD-H})
is needed. 
The relevant series expansion of the auxiliary
functions~\eqref{f1-H}--\eqref{f3-H} (near fields) reads as
\begin{align}
\label{f0-H-ser}
f_0(R,\ell)&=\frac{1}{\ell}\, R-\frac{1}{2 \ell^2}\, R^2
+\mathcal{O}(R^3)\,,\\
\label{f1-H-ser}
f_1(R,\ell)&=\frac{1}{2\ell^2}\, R^2-\frac{1}{3 \ell^3}\, R^3+\frac{1}{8 \ell^4}\, R^4+\mathcal{O}(R^5)\,,\\
\label{f2-H-ser}
f_2(R,\ell)&=\frac{1}{6\ell^2}\,R^2-\frac{1}{24 \ell^4}\, R^4+\mathcal{O}(R^5)\,,\\
\label{f3-H-ser}
f_3(R,\ell)&=\frac{1}{10\ell^2}\,R^2-\frac{1}{120\ell^4}\, R^4+\mathcal{O}(R^5)\,.
\end{align}
From Eq.~\eqref{f0-H-ser}
it is obvious that the function $f_0(R,\ell)$ regularizes up to a
$1/R$-singularity and gives a non-singular field expression.
From Eqs.~\eqref{f1-H-ser}--\eqref{f3-H-ser}
it can be seen that the functions $f_1(R,\ell)$,  $f_2(R,\ell)$
and  $f_3(R,\ell)$ regularize up to a $1/R^2$-singularity and give
non-singular fields. 
At $R=0$ the  auxiliary functions~(\ref{f0-H})--(\ref{f3-H}) are zero and 
in the far field they approach 1.

Thus, the displacement field is non-singular and finite at $R=0$, but suffers a jump
due to $\tau_i f_1(R,\ell)/R^2$.
The first displacement gradient possesses a $1/R$-singularity, and 
the second displacement gradient possesses a $1/R^2$-singularity
(see Fig.~\ref{fig:f}).

Using the 3D Green function of the Helmholtz equation, $G^\text{H}(R)$, with 
$G^\text{H}(R)*1/R^n$,
gradient elasticity of Helmholtz type regularizes singularities up to 
$1/R^2$, namely  $G^\text{BH}(R)*1/R^2$ is finite. 
For the order $n=3$, $G^\text{BH}(R)*1/R^3\sim 1/R$,
and for the order $n=4$, $G^\text{BH}(R)*1/R^4\sim 1/R^2$.

Moreover,
it is interesting to notice that
Eqs.~(\ref{u-PD-H})--(\ref{u-grad2-PD-H})  agree with the expressions given in~\cite{Dobov06}.

\subsection{Limit to classical elasticity}

The limit from gradient elasticity of Helmholtz type to classical elasticity is $\ell\rightarrow 0$ and we obtain from 
Eqs.~(\ref{u-PD-H}) and (\ref{u-grad-PD-H}) 
the classical displacement field of a dilatation centre 
\begin{align}
\label{u-PD-0}
u^0_i=-\frac{Q}{4 \pi} \frac{1+\nu}{1-\nu}\, \pd_i \bigg(\frac{1}{R}\bigg)
=\frac{Q}{4 \pi} \frac{1+\nu}{1-\nu}\, \frac{R_i}{R^3}\,
\end{align}
and its gradient, being the total distortion tensor, 
\begin{align}
\label{u-grad-PD-0-2}
u^0_{i,j}
&=-\frac{Q}{4 \pi} \frac{1+\nu}{1-\nu}\, \pd_j\pd_i \bigg(\frac{1}{R}\bigg)\nonumber\\
&=\frac{Q}{4 \pi} \frac{1+\nu}{1-\nu} \bigg[\frac{\delta_{ij}}{R^3}\,
 -\frac{3R_i R_j}{R^5}
+\frac{4\pi}{3}\,   \delta_{ij}\,  \delta(\BR) 
 \bigg]\,.
\end{align}
The total dilatation, which follows from Eq.~(\ref{u-grad-PD-0-2}), 
\begin{align}
\label{u-div-PD-0}
u^0_{i,i}=Q\, \frac{1+\nu}{1-\nu} \, \delta(\bm R)
\end{align}
and the quasi-plastic dilatation of a dilatation centre 
\begin{align}
\label{BP-0}
\beta^{\TP,0}_{ij}=Q\, \delta_{ij} \delta(\BR)
\end{align}
are given in terms of a Dirac delta function. 
The displacement field possesses a $1/R$-singularity, and 
the displacement gradient possesses a $1/R^2$-singularity
and a $\delta(\BR)$-singularity. 
Note that the Dirac delta term in the total distortion~(\ref{u-grad-PD-0-2})
is often erroneously neglected (see, e.g., \cite{Teodosiu}).

\section{Interaction energy and interaction force between defects in gradient elasticity of bi-Helmholtz type}
\label{Force}

In this Section, we investigate the interaction between two dilatation centres as well as a dilatation centre with an edge dislocation
in the framework of gradient elasticity of bi-Helmholtz type. 
We consider case~(1) as characteristic example. 
As we have shown in Section~\ref{PD}, cases~(2) and (3) have a similar
behaviour as case~(1) and 
can be easily obtained from case~(1).


\subsection{Interaction energy}

In gradient elasticity of bi-Helmholtz type, the interaction energy is given by 
\begin{align}
\label{U-int-0}
U_{\text{int}}=\int_V  
\big( \sigma_{ij}e_{ij}
+\tau_{ijk} \pd_k e_{ij}
 +\tau_{ijkl} \pd_l \pd_k e_{ij}\big)
 \, \d V
=\int_V L \sigma_{ij} e_{ij}\, \d V
=-\int_V \sigma_{ij} \beta^{\TP,0}_{ij}\, \d V\,,
\end{align}
where we have used integration by parts, the Green-Gauss theorem with vanishing surface terms 
at infinity and Eqs.~(\ref{uIJ}) and (\ref{BP-H-RA}).

\subsubsection{Interaction energy between two point defects}

Substituting Eq.~(\ref{BP-PD-0}) into Eq.~(\ref{U-int-0}), 
the interaction energy between a point defect 
with $Q_{ij}$ in the stress field $\sigma_{ij}$ of another defect, or 
with $P_{ij}$ in the elastic strain field $e_{ij}$,
reads
(see, e.g.,~\cite{Lazar17})
\begin{align}
\label{U-int-1}
U_{\text{int}}=-Q_{ij}\sigma_{ij}=-P_{ij} e_{ij}\,.
\end{align}
For a dilatation centre, the interaction energy~(\ref{U-int-1}) reduces to
 \begin{align}
\label{U-int-2}
U_{\text{int}}=-Q\sigma_{ii}=-P e_{ii}\,.
\end{align}
Substituting Eqs.~(\ref{dil}) and (\ref{P}) into 
(\ref{U-int-2}), the interaction energy becomes
\begin{align}
\label{U-int-3}
U_{\text{int}}
=Q Q'\, 4\mu\, \frac{1+\nu}{1-\nu}\, G^{\text{BH}}(R)\,.
\end{align}
Using the Green function~(\ref{G-BH}), 
the interaction energy~(\ref{U-int-2}) between the two dilatational centres 
reduces to 
\begin{align}
\label{U-int-4}
U_{\text{int}}
=\frac{Q Q'\mu (1+\nu)}{\pi(1-\nu)}\, 
\frac{1}{(c_1^2-c_2^2)}\,\frac{1}{R} 
\Big(\e^{-R/c_1}-\e^{-R/c_2}\Big)\,.
\end{align}
In Fig.~\ref{fig:GF}(a), it can be seen that the interaction energy~(\ref{U-int-3}) is 
a short-range interaction energy  and remains finite if $R\rightarrow 0$. 
For $QQ'<0$, the interaction energy~(\ref{U-int-4}) possesses a minimum at
$R=0$, and 
for $QQ'>0$, the interaction energy~(\ref{U-int-4}) possesses a maximum at
$R=0$.
In fact, using gradient elasticity of bi-Helmholtz type, 
the finite short-range interaction energy is the regularization 
of the contact term in the interaction energy discussed in~\cite{Lazar17}:
\begin{align}
\label{U-int-cl}
\lim_{c_1\to 0,\,  c_2 \to 0}
U_{\text{int}}
=Q Q'\, 4\mu\, \frac{1+\nu}{1-\nu}\, \delta(\BR)\,.
\end{align}

In the limit to gradient elasticity of Helmholtz type, $c_1\rightarrow\ell$ and $c_2\rightarrow 0$, the interaction energy~(\ref{U-int-4})  reduces to 
\begin{align}
\label{U-int-Y}
U_{\text{int}}
=\frac{Q Q'\mu (1+\nu)}{\pi(1-\nu)\ell^2}\, 
\,\frac{\e^{-R/\ell}}{R} \,,
\end{align}
which has the form of a Yukawa potential known from the physics of nucleons \cite{Wentzel,Felsager}.
For $R\rightarrow 0$, the Yukawa potential has a $1/R$-singularity.
Therefore, the non-singular interaction energy~(\ref{U-int-4}) between two dilatation centres may be called bi-Yukawa potential. 

\subsubsection{Self-energy of a dilatation centre}

Using Eq.~(\ref{G-BH-0}), Eq.~(\ref{U-int-3}) delivers the  finite self-energy of a dilatation centre with $Q=Q'$
\begin{align}
\label{U-s}
U_{\text{s}}=\frac{1}{2}\, U_{\text{int}}(0)
=\frac{Q^2}{2\pi}\, \mu\, \frac{1+\nu}{1-\nu}\,
\frac{1}{c_1 c_2(c_1+c_2)}\,.
\end{align}
Eq.~(\ref{U-s}) means that the self-energy of a point defect is equivalent to half the interaction energy between two identical point 
defects when $R=0$. 
The finite Green function (form factor) (\ref{G-BH}) has assured a finite self-energy.

\subsection{Interaction force}

The interaction force can be written as (negative) gradient of the interaction energy (see also \cite{Lazar17})
\begin{align}
\label{PK-1}
\FF_k=-\pd_k U_{\text{int}}\,. 
\end{align}

\subsubsection{Interaction force between two point defects}

Using the interaction energy~(\ref{U-int-1}), 
we obtain from Eq.~(\ref{PK-1}) the force exerted on the point defect 
in the gradient of a stress field 
or in the gradient of an elastic strain field
\begin{align}
\label{PK-2}
\FF_k
=Q_{ij} \sigma_{ij,k}=P_{ij} e_{ij,k}\,.
\end{align}
Eq.~(\ref{PK-2}) gives
the interaction force between a point defect
of strength $Q_{ij}$ and the stress gradient field $\sigma_{ij,k}$,  
which can be caused by other defects (point defect, dislocation). 
On the other hand, the material force~(\ref{PK-2})
can be expressed in terms of the elastic dipole tensor $P_{ij}$ and 
the elastic strain gradient tensor 
(see also \cite{Kroener58,Kroener60,Lazar17}). 
For a dilatation centre, the interaction force~(\ref{PK-2}) simplifies to 
\begin{align}
\label{PK-3}
\FF_k
=Q \sigma_{ii,k}=P e_{ii,k}\,,
\end{align}
where
\begin{align}
\label{P}
P=2\mu\, \frac{1+\nu}{1-2\nu}\, Q\,.
\end{align}
Using Eqs.~(\ref{dil}) and (\ref{P}), we obtain from Eq.~(\ref{PK-3})
\begin{align}
\label{PK-4}
\FF_k
=-  Q Q'\, 4\mu\, \frac{1+\nu}{1-\nu}\, \pd_k G^{\text{BH}}(R)\,.
\end{align}
Substituting Eq.~(\ref{G-grad-BH}) into (\ref{PK-4}), 
the interaction force between two dilatation centres reads as 
\begin{align}
\label{PK-5}
\FF_k
=  \frac{Q Q' \mu\, (1+\nu)}{\pi(1-\nu)}\, 
\frac{1}{(c_1^2-c_2^2)}\, 
\frac{R_k}{R^3 }\bigg[\e^{-R/c_1}-\e^{-R/c_2}
+\frac{R}{c_1}\, \e^{-R/c_1}-\frac{R}{c_2}\,\e^{-R/c_2}
\bigg]\,.
\end{align}
Therefore, the interaction force~(\ref{PK-5}) between two dilatation centres  is a 
short-range interaction force.  
The force~(\ref{PK-5}) is the short-range interaction force exerted by one point defect 
with strength $Q'$ at $\Bx'$ on the other point defect with strength 
$Q$ at $\Bx$. 
Here, $R$ is the distance between the two defects from $Q'$ to $Q$, 
and $\BR=\Bx-\Bx'$ is the vector from $Q'$ to $Q$.
Using Eq.~(\ref{G-grad-BH-0}), it can be seen that Eq.~(\ref{PK-4}) is finite at $R=0$
\begin{align}
\label{PK-max}
\FF_k(0)
=\frac{Q Q'}{2\pi}\, \mu\, \frac{1+\nu}{1-\nu}\, \frac{1}{c_1^2 c_2^2}\, \tau_k\,.
\end{align}

In the limit to gradient elasticity of Helmholtz type, $c_1\rightarrow\ell$ and $c_2\rightarrow 0$, the interaction force~(\ref{PK-5})  reduces to a singular Yukawa type force
\begin{align}
\label{PK-6}
\FF_k
=  \frac{Q Q' \mu\, (1+\nu)}{\pi(1-\nu)\ell^2}\, 
\frac{R_k}{R^3 }\bigg(1+\frac{R}{\ell}\bigg)\,\e^{-R/\ell}\,.
\end{align}
Therefore, the non-singular interaction force~(\ref{PK-5}) between two dilatation centres is a bi-Yukawa type force.

\subsection{Interaction between an edge dislocation and a dilatation centre}

The aim of this subsection is to study the interaction between an edge dislocation and a point defect (dilatation centre)
which is of importance for the understanding of the properties of solids.
 We consider a point defect in the stress field of an edge dislocation. 
 
The hydrostatic stress field of an edge dislocation with (positive) Burgers vector $b_x$ located at $(x,y)=(0,0)$ 
in the framework of gradient elasticity of bi-Helmholtz type is given by~\cite{LMA06}
\begin{align}
\label{T-ii}
\sigma_{ii}=-\frac{\mu b_x(1+\nu)}{\pi(1-\nu)}\,\frac{\sin\varphi}{r}
\Big\{1-\frac{1}{c_1^2-c_2^2}\big[c_1 r K_1( r/c_1)-c_2 r K_1(r/c_2)\big]\Big\}\,,
\end{align}
where $r=\sqrt{x^2+y^2}$.

\subsubsection{Interaction energy}

\begin{figure}[t]\unitlength1cm
\vspace*{0.1cm}
\centerline{
\epsfig{figure=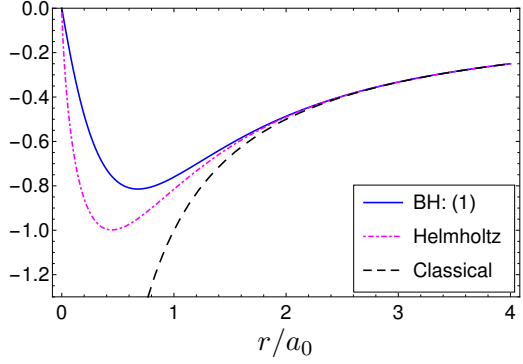,width=6.8cm}
\put(-3.5,-0.4){$r/a_0$}
}
\caption{Interaction energy between a point defect and an edge dislocation
in gradient elasticity of bi-Helmholtz type,
gradient elasticity of Helmholtz type
and classical elasticity
for  $\varphi=3\pi/4$ in units of $\mu Q b_x(1+\nu)/[\pi(1-\nu)a_0]$.}
\label{fig:U}
\end{figure}

If Eq.~(\ref{T-ii}) is substituted into (\ref{U-int-2}), 
the interaction energy between an edge dislocation and a dilatation centre is  obtained as 
\begin{align}
\label{U-PD-D}
U_{\text{int}}=\frac{\mu Q b_x(1+\nu)}{\pi(1-\nu)}\,\frac{\sin\varphi}{r}
\Big\{1-\frac{1}{c_1^2-c_2^2}\big[c_1 r K_1( r/c_1)-c_2 r K_1(r/c_2)\big]\Big\}\,.
\end{align}
The point defect is located at $(r,\varphi)$. 
For an oversized point defect ($Q>0$), $U_{\text{int}}$ is positive for positions above the slip plane ($0<\varphi<\pi$) 
and negative below  ($\pi<\varphi<2\pi$) since the edge dislocation produces compression in the region of the extra half-plane 
and tension below.  For an undersized point defect ($Q<0$) the positions of attraction and repulsion are reversed. 
An important feature of this result is that, contrary to the classical result,
the interaction energy remains finite if $r\rightarrow 0$ 
as it can be seen in Fig.~\ref{fig:U}. 
The minimum of the interaction energy near the dislocation line at $\varphi=\pi/4$ 
gives the binding energy between an  edge dislocation and a vacancy (negative dilatation centre)  or undersized poind defect 
($Q<0$) and at $\varphi=3\pi/4$ 
gives the binding energy between an  edge dislocation and an oversized point defect (positive dilatation centre) ($Q>0$).
The position and the value of the binding energy depend on the lengths $c_1$ and $c_2$.  
If $c_2\rightarrow 0$ and $c_1\rightarrow 0$, 
then the classical Cottrell-Bilby result is recovered~\cite{CB}.

In the limit to gradient elasticity of Helmholtz type, $c_1\rightarrow\ell$ and $c_2\rightarrow 0$, the interaction energy~(\ref{U-PD-D})  reduces to 
\begin{align}
\label{U-PD-D-H}
U_{\text{int}}=\frac{\mu Q b_x(1+\nu)}{\pi(1-\nu)}\,\frac{\sin\varphi}{r}
\Big\{1-\frac{r}{\ell}\,K_1( r/\ell)\Big\}\,,
\end{align}
which is still singularity-free (see Fig.~\ref{fig:U}). 
For $Q>0$ at $\varphi=3\pi/4$, the interaction energy~(\ref{U-PD-D-H}) 
has minimum of $U_{\text{int}}=0.399 \mu Q b_x(1+\nu)/[\pi(1-\nu)\ell]$ at 
$r=1.114\ell$. 
Using the value of $\ell=0.49 a_0$ 
for Al ($a_0=4.05$\,\AA) calculated in the framework of gradient
elasticity of Helmholtz type~\cite{Lazar17b,Shodja13},
the position of the minimum of the interaction energy (binding energy)
is $U_{\text{int}}=0.814 \mu Q b_x(1+\nu)/[\pi(1-\nu)a]$ 
at $r= 0.546 a_0=0.772 b_x$ where $b_x=a_0/\sqrt{2}$ is the Burgers vector in fcc.  
Moreover, it is interesting to notice that 
Eq.~(\ref{U-PD-D-H}) is in agreement with the expression given in~\cite{Vlasov}.

\subsubsection{Interaction force}

\begin{figure}[t]\unitlength1cm
\vspace*{0.1cm}
\centerline{
\epsfig{figure=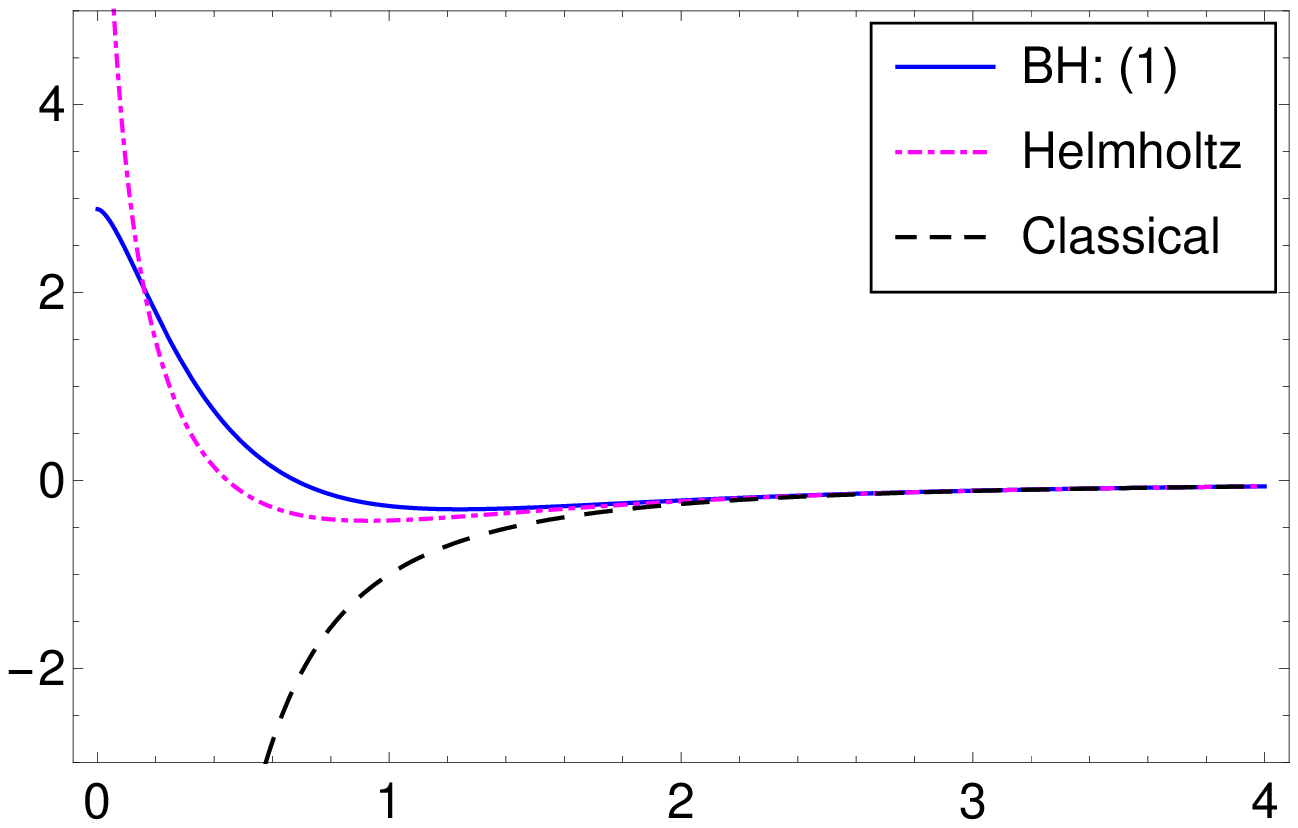,width=6.6cm}
\put(-3.5,-0.4){$r/a_0$}
\put(-6.5,-0.4){$\text{(a)}$}
\hspace*{0.4cm}
\put(0,-0.4){$\text{(b)}$}
\put(3.0,-0.4){$r/a_0$}
\epsfig{figure=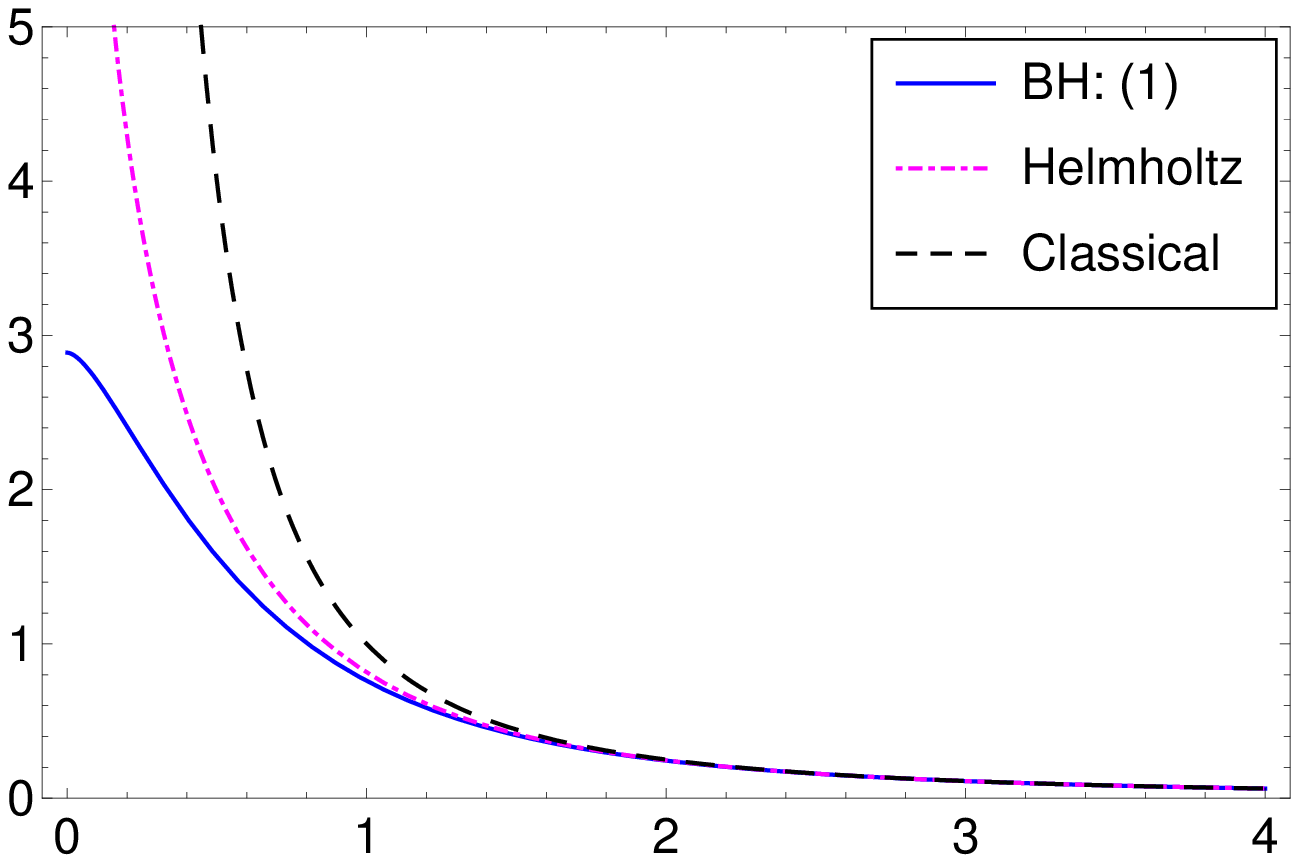,width=6.4cm}
}
\caption{Interaction force between a point defect and an edge dislocation
in gradient elasticity of bi-Helmholtz type,
gradient elasticity of Helmholtz type and classical elasticity:
(a) $\FF_r$,
(b) $\FF_\varphi$,
for $\varphi=3\pi/4$ 
in units of $\mu Q b_x(1+\nu)/[\pi(1-\nu)a_0^2]$.}
\label{fig:PKF}
\end{figure}

The interaction force between 
an edge dislocation and a dilatation centre is given by 
\begin{align} 
\label{PK-PD-D-r}
\FF_r&=-\frac{\pd U_{\text{int}}}{\pd r}\nonumber\\
&=
\frac{\mu Q b_x(1+\nu)}{\pi(1-\nu)}\,
\frac{\sin\varphi}{r^2}\Big\{
1-\frac{1}{c_1^2-c_2^2}
\big[c_1 r K_1(r/c_1)-c_2 r K_1(r/c_2)\big]
\nonumber\\
&\hspace{5cm}
-\frac{r^2}{c_1^2-c_2^2}
\big[K_0(r/c_1)-K_0(r/c_2)\big]\Big\}
\end{align}
and 
\begin{align}
\label{PK-PD-D-p}
\FF_\varphi&=-\frac{1}{r}\,\frac{\pd U_{\text{int}}}{\pd \varphi}\nonumber\\
&=-\frac{\mu Q b_x(1+\nu)}{\pi(1-\nu)}\,\frac{\cos\varphi}{r^2}
\Big\{1-\frac{1}{c_1^2-c_2^2}\big[c_1 r K_1( r/c_1)-c_2 r K_1(r/c_2)\big]\Big\}
\,.
\end{align}
Both components $\FF_r$ and $\FF_\varphi$ are plotted in Fig.~\ref{fig:PKF} and it can be seen that they
are finite at $r=0$. 
Figure~\ref{fig:PKF} shows the interaction force as a function of the distance between the
edge dislocation and the point defect. 
It can be seen in Fig.~\ref{fig:PKF}(a)  that near the edge dislocation  
a bound state appears when $\FF_r=0$ (position of minimum of
$U_{\text{int}}$). 
The position at $\FF_r=0$ is the equilibrium position of 
the point defect in the stress field of an edge dislocation
not present in classical elasticity. 
We notice a significant deviation between the classical results and the gradient results  
in Figs.~\ref{fig:U} and \ref{fig:PKF}
when the defect separation is less than $r=2 a_0$ because classical elasticity
is not valid at such a scale unlike gradient elasticity. 
In particular, $\FF_r$ changes the sign near the dislocation unlike the
classical result (see Fig.~\ref{fig:PKF}(a)).

In the limit to gradient elasticity of Helmholtz type, $c_1\rightarrow\ell$ and $c_2\rightarrow 0$, 
the interaction force between an edge dislocation and a dilatation centre,
Eqs.~(\ref{PK-PD-D-r}) and (\ref{PK-PD-D-p}), reduces to 
\begin{align} 
\label{PK-PD-D-r-H}
\FF_r&=
\frac{\mu Q b_x(1+\nu)}{\pi(1-\nu)}\,
\frac{\sin\varphi}{r^2}\Big\{
1-\frac{r}{\ell}\, K_1(r/\ell)
-\frac{r^2}{\ell^2}\,K_0(r/\ell)\Big\}
\end{align}
and 
\begin{align}
\label{PK-PD-D-p-H}
\FF_\varphi
&=-\frac{\mu Q b_x(1+\nu)}{\pi(1-\nu)}\,\frac{\cos\varphi}{r^2}
\Big\{1-\frac{r}{\ell}\, K_1( r/\ell)\Big\}
\,.
\end{align}
$\FF_r=0$ at $r=1.114\ell$.
Eqs.~(\ref{PK-PD-D-r-H}) and (\ref{PK-PD-D-p-H})  possess a $\ln r$-singularity at $r=0$.  
Using the value of $\ell=0.49 a_0$ 
for Al calculated in the framework of gradient
elasticity of Helmholtz type~\cite{Lazar17b,Shodja13},
the equilibrium position of the point defect in the stress field of an edge dislocation 
reads: $r= 0.546 a_0$. 
For Aluminum, it reads: $r=2.211\, \text{\AA}$. 
Note that Eq.~(\ref{PK-PD-D-r-H}) is in agreement with the expression given in~\cite{Vlasov}.

\section{Conclusion}
\label{Concl}
In this paper, we have presented and developed a non-singular continuum theory of point defects.
It is based on a second strain gradient elasticity theory, 
the so-called gradient elasticity of bi-Helmholtz type, 
with eigenstrain caused by point defects.
We discussed possible values for the two characteristic internal lengths of gradient elasticity of bi-Helmholtz type ($\ell_1$, $\ell_2$),
which are two material parameters describing the range of the weak nonlocality present near the defects.  
Our model is able to give non-singular expressions for the displacement field, the first displacement gradient, the second 
displacement gradient, the plastic distortion and the first gradient of the plastic distortion. 
This fact makes it possible to solve a set of problems unsolvable in classical elasticity as well as in 
gradient elasticity of Helmholtz type.
Unlike the singular expression in classical elasticity and in gradient elasticity of Helmholtz type, 
we have found non-singular and finite expressions for:
\begin{itemize}
\item
interaction energy between two dilatation centres (bi-Yukawa potential)
\item
interaction force between two dilatation centres (bi-Yukawa force)
\item
self-energy of a dilatation centre
\item
interaction force between a dilatation centre and an edge dislocation\,.
\end{itemize}
Moreover, unlike the singular expression in classical elasticity, 
we have found
\begin{itemize}
\item
interaction energy between a dilatation centre and an edge dislocation
\end{itemize}
similar to the non-singular expression obtained in gradient elasticity of
Helmholtz type leading to realistic expressions of binding energies and of the
equilibrium position of a point defect in the stress field 
of an edge dislocation. 
The obtained fields of point defects are non-singular due to a 
straightforward and self-consistent regularization 
based on the bi-Helmholtz operator and Green function.
The non-singular fields are important and necessary 
for a better discrete dislocation dynamics and
computer simulations of non-singular defects.
Moreover, all the obtained non-singular analytical expressions
can be compared with corresponding atomistic and experimental results.

For the three-dimensional solutions (e.g. Green functions, displacement and displacement gradients)
in gradient elasticity of bi-Helmholtz type, there are three types of solutions:
\begin{itemize}
\item
in case~(1), the solutions are given in terms of two decreasing exponential functions $\exp(-r/c_1)$ and $\exp(-r/c_2)$
\item
in case~(2), the solutions are given in terms of one decreasing exponential function $\exp(-r/c_1)$
\item
in case~(3), the solutions are given in terms of one decreasing exponential function $\exp(-a r)$ 
times sine and/or cosine functions: $\sin(br)$ and $\cos(br)$.
\end{itemize}

Although the auxiliary lengths $c_1$ and $c_2$ can be 
real or complex conjugate, 
the material lengths $\ell_1$ and $\ell_2$ are always
real. 
In other words, 
independent if the auxiliary lengths $c_1$ and $c_2$ are real or 
complex conjugate, the physical behaviour of the displacement and displacement gradients of a dilatation centre is not changed and 
it does not show oscillations.  Therefore, the physical behaviour  of the displacement and displacement gradients of a dilatation centre 
in the possible cases~(1), (2) and (3) is similar.

\section*{Acknowledgements}
The author gratefully acknowledges a grant from the
Deutsche Forschungsgemeinschaft
(Grant No. La1974/4-1).

\end{document}